\documentclass[preprint,amsmath,amssymb
]{revtex4-1}
\usepackage[T1]{fontenc}
\usepackage{graphicx}
\usepackage{bm}
\usepackage{mathtools}
\usepackage{epsfig} 
\usepackage{subfigure}
\usepackage{url}
\usepackage{amsmath}
\usepackage[dvipsnames]{xcolor}
\usepackage{comment}
\usepackage{gensymb}
\usepackage{float}

\definecolor{darkblue}{rgb}{0.00,0.00,0.50}
\definecolor{darkgreen}{rgb}{0.00,0.50,0.00}
\definecolor{violet}{rgb}{0.4,0,0.3}
\definecolor{gris}{gray}{0.85}

\begin{document}

\title{Geometric theory of topological defects: methodological developments and new trends}

\author{S\'ebastien Fumeron$^\ddag$}
\author{Bertrand Berche$^\ddag$}
\author{Fernando Moraes$^\dagger$}
\affiliation{$^\ddag$Laboratoire de Physique et Chimie Th\'eoriques, UMR Universit\'e de Lorraine - CNRS 7019, 54000  Nancy, France}
\affiliation{$^\dagger$Departamento de F\'{\i}sica, Universidade Federal Rural de Pernambuco, 52171-900, Recife, PE, Brazil}

\begin{abstract}
Liquid crystals generally support orientational singularities of the director field known as topological defects. These latter modifiy transport properties in their vicinity as if the geometry was non-Euclidean. We present a state of the art of the differential geometry of nematic liquid crystals, with a special emphasis on linear defects. We then discuss unexpected but deep connections with cosmology and high-energy-physics, and conclude with a review on defect engineering for transport phenomena.  
\end{abstract}

\maketitle

\section{Introduction}

\textcolor{black}{One of Pierre-Gilles de Gennes's greatest breakthrough was to realize that methods and concepts borrowed from superconductivity also apply to describe smectic-A phases \cite{de1972analogy}. His work is a striking example of cross-fertilization between different areas of physics and it highlights how progress arises at the crossroads of various scientific fields. In an article that has not been translated in English \cite{gennes1989cinquante}, he took the example of line singularities as a common denominator between liquid crystals, quark physics and superconductors \footnote{The cases of disclination lines in nematics, gluon flux tubes and vortices in superconductors are discussed pp. 48-49 of \cite{gennes1989cinquante}, where he wrote: ``I found quite extraordinary the deep relationship there is between molecular liquids such as nematics, at one end, and the building block of matter at the subnuclear scale.``}. The same observation was also made by William Brinkman and Patricia Cladis \cite{cladis1982defects}, and most notably by the two-Nobel-Prize winner John Bardeen in a review written as a plea for interdisciplinarity: ``\textit{Line defects in three-dimensional systems, quantized vortex lines or flux lines, and dislocations account for similarities of behavior in superconductors, liquid crystals, and, it is hoped, color confinement of quarks}``\cite{bardeen1980unity}.}

\textcolor{black}{In this spirit, the objective of this paper is to perform an in-depth survey of geometrical methods useful for investigating topological defects and to describe some of its modern applications, either as a playground to test fundamental ideas in high-energy physics or gravitational physics, or as high-performance tools to taylor transport phenomena from soft matter devices. We will be mainly concerned with nematic liquid crystals and topological line defects.}

\textcolor{black}{In section II, we provide a self-contained introduction to phase transitions, geometry, topology and optics in such systems. We show how a metric description of defect lines in terms of Riemann manifolds naturally arises in nematics, before addressing the question of analogue gravity which may be less familiar to the liquid crystals community.}

\textcolor{black}{Section III is an introduction to some of the ideas borrowed to cosmology which can be dealt with liquid crystals, such as the Kibble mechanism, which rules the formation of defects in the early universe but also in nematics. We review several outstanding problems involving line singularities, such as cosmic strings, wormholes and bouncing cosmologies, and discuss their connections with disclinations in liquid crystals.} 

Section IV eventually discusses applications of the geometric formalism previously introduced for the description of acoustics, optics and heat transfer in the presence of such defects. \textcolor{black}{The main idea is to show how the curvature carried by the topological defects can be used to design specific propagation patterns, a possible step forward towards the defect-engineering of transport phenomena.}  

\section{Topological defects in nematic liquid crystals}

\subsection{Basics of the isotropic-nematic phase transition}

\paragraph{Historical milestones} The story of liquid crystal has met with a shaky start. The first observations reported of what is today understood as a liquid crystal belong to the realm of biology. Georges-Louis Buffon (1707-1788) and later Rudolf Virchow (1854) and Carl Mettenheimer (1855) reported about the strange behavior of lecithins, a family of phospholipid substances contained in plants (wheat, rye...) and in animals (yolks, myelin - the insulating coating of nerve fibres - ...) \cite{palffy2007histo}. When suspended in water, lecithins form birefringent tubular structures like a Iceland spar but that writhed like eels. Julius Planer in 1861 \cite{lisetski2010observed} and most importantly Friedrich Reinitzer in 1888 discovered similar optical behaviors with cholesterol compounds. Reinitzer extracted cholesteryl esters from carrots and made an unanticipated observation: contrary to what was known in crystallography, cholesteryl benzoate displays two melting points \cite{oswald2005nematic}. The lower one occurs at about $145.5\degree$C and correspond to the melting of the solid phase into a turbid fluid. The higher melting point corresponds to the clarification of the milky liquid beyond $178.5 \degree $C.          

Such behavior left Reinitzer skeptical: had he discovered a genuinely new behavior of matter or was this simply the result of impure and incompletely melted crystals? Reinitzer wrote to Otto Lehmann, a leading physicist known for designing the first ``crystallization microscope'': this latter consists in a microscope equipped with crossed polarizers and a thermal deck (a small Bunsen burner and two cooling blasts) to observe how crystals behave when the temperature varies. Lehmann reproduced and improved Reinitzer's observations, and promoted these substances as new forms of matter, half-between liquids and crystals. In 1889, he coined the term ``liquid crystals'' to account for his discovery (somehow pulling the sheet back towards him as he even claimed priority over Reinitzer \cite{mitov2014liquid}). Soon afterwards, he kept changing its name (including ``flowing crystals'', ``crystalline liquids''...), which reveals his difficulties to grasp the real nature of what he found. 

The years that followed Lehmann's breakthrough have been critical. On one hand, the subject became more and more discussed in the scientific community, even drawing the attention of future Nobel Prize laureates such as Max Born (in 1916, he conceived the first molecular theory of liquid crystals but predicted a generic ferroelectric behavior that turned out to be incorrect \cite{born1916anisotropic}), Jacobus Henricus van’t Hoff and Walther Nernst (Lehmann himself was an unlucky nominee from 1913 to 1922). On the other hand, the subject became highly controversial: partly because Nernst and most especially Gustav Tammann led a vivid opposition against liquid crystals (suspecting them of being nothing more than poorly prepared colloidal mixtures), partly because of Lehmann's personality (a sparkling mix of pretentiousness and mysticism \cite{mitov2014liquid}).

Liquid crystals have stayed a controversial subject, since the decisive contributions of Rudolf Schenk (1905), Daniel Vorländer (1907) and George Friedel (1922) to get a clear view on this subject. Schenk led a thorough study of the clearing point and unambiguously showed the observed properties (density, viscosity) could not be related to mixtures. Vorländer elucidated the mysterious anisotropic behavior exibited by the fluid: from a microscopic standpoint, liquid crystals consist in self-organized assemblies of rod-like molecules. As such, they exibit both the birefringence property expected from anisotropic uniaxial media and the ability to flow. Finally, Friedel realized that such substances should properly be understood as new full-blown phases of matter, that he named mesophases. Initially divided into three broad families (nematic, cholesteric and smectic), liquid crystals have now been enriched by many new mesophases, including columnar phases, cubatic phases, blue phases I, II and III...    

\paragraph{Mesogenic behavior} The recipe for a molecule to be nematogenic (i.e. to have the ability to organize into a nematic phase) is rather simple: take a rod-like molecule and deck it with 1) a flexible outer part (an aliphatic chain), 2) a rigid core (phenyl groups most generally) and 3) a chemical group bearing a permanent dipole (for instance, a carbonitrile group as in 5CB). The resulting substance is a thermotropic nematic, a sensitive compromise between the attractive Van der Waals interactions that align rigid cores on average along the same direction (anisotropy) and the thermal agitation of the aliphatic chains increasing the mean steric hindrance (fluidity). Nematics can also be lyotropic: the nematogens display an amphiphilic structure (they have both hydrophilic and hydrophobic parts), the control parameter being the concentration of  molecules in a solvent such as water. The frontier between thermotropic and lyotropic liquid crystalline behavior being not strict, nematogens can also behave as amphotropic media.

In the perspective of section III, let us focus on thermotropic nematics. In that case, there are different models accounting adequately for the phase transition. The Lebwohl-Lasher model \cite{lebwohl1972nematic} is the paradigmatic model in this context, in a sense the liquid crystal analogue of the Ising model \cite{fate}: the nematogen molecules are represented only by their direction and they occupy fixed positions on the sites of a cubic lattice. The different sites of the lattice interact only between nearest neighbors through a potential that favors configurations where the neighboring molecules point in the same direction. On the contrary, the Maier-Saupe approach is a mean-field theory: interactions between particles are replaced by an effective field experienced by all particles at the same time. This model considers only London forces between instantaneous molecular dipoles and ignores repulsive interactions \cite{maier1958einfache}. This also favors alignment of the molecules in the same direction. We will briefly mention that for lyotropic nematics, Lars Onsager described the phase transition of an assembly of rod-like sticks as an entropic process driven \cite{onsager1949effects}: due to purely steric effects, orientational entropy loss is more than offset by positional entropy gain which triggers the transition.  

Depending on the temperature range, three (or more) phases can  be observed. At low temperatures, the steric hindrance of aliphatic chains is minimal and the nematogens get close enough for attractive forces to drive the system. As the dipole-dipole interactions prevail over thermal agitation, the assembly of rod-like molecules organize into a molecular crystal. This latter displays both a positional and rotational order for each molecule. On the contrary, at high temperatures, Van der Waals interactions are dominated by thermal effects and the nematogens form an isotropic fluid phase: both positional and orientational order are lost. Within the intermediate range of temperature, the two effects are of the same order and  different kinds of mesophases may appear. In the nematic mesophase, only the orientational order is preserved: locally, the nematogens tend to align on average along a common direction, which defines the director field $\mathbf{n}$. In usual nematics, the orientational order is preserved at long distance, as the correlation length is typically about a few $\mu$m, compared to the nematogen length around a few nanometers. As the phase transition involves nucleation, these domains wherein nematogens share a common orientation form submicronic bubbles (or spherulites). Then they grow in size and eventually they meet and mingle, sometimes leaving relics in the form of long threads.

\paragraph{Order parameter} Within a nematic, a particular molecule  does generally not point exactly  in the direction $\mathbf{n}$ and the degree of orientational ordering of the mesophase can thus be assessed by looking how well the nematogens are aligned along the director field. The quadrupolar scalar order parameter $S$ defined by Tsvetkov (1942) provides a quantitative criterion to characterize the nematic order: it is normalized ($S=0$ in the isotropic fluid phase and $S=1$ for perfectly aligned rods), and ranges from $0.3<S<0.8$ in usual nematics (in practice, $0.3<S<0.4$ for thermotropic liquid crystals, whereas $0.6<S<0.8$ for lyotropic ones). The order parameter can be refined to include informations on the local orientation of $\mathbf{n}$ (Landau -- de Gennes tensorial order parameter) or to encompass phase involving more complex-shaped mesogens (higher-order mutipole-multipole correlation functions).

For thermotropic nematics, $S$ can be taken as a function depending only on the temperature. The behavior of $S$ at the transition can essentially discriminate between two main families of phase transitions (in the sense defined by Lev Landau in 1937 \cite{landauPhyStat}):  first-order phase transitions, for which the order parameter displays a jump at the transition  control parameter (this class also involves latent heats and nucleation processes), and  second-order phase transitions, for which the order parameter varies continuously at the transition (this class involves pretransitional effects and scaling behaviors). Experimentally, for most compounds (5CB, MBBA, 5CN...), the isotropic-nematic phase transition is identified as weakly first-order phase transition in three dimensions. It combines small discontinuities of $S$ \cite{Chandrasekhar1992}, nucleation \cite{kleman2006topological} and low latent heats \cite{van2005weakly}, but pretransitional effects of the dielectric properties \cite{van2005weakly}.  

\subsection{From symmetry to topology} 

\paragraph{Homotopy theory} The existence of orientational and/or positional orders impart each phase about the transition with a specific set of symmetries. As a rule, the higher-temperature phase is generally the less-ordered one and its symmetry group is larger. In the isotropic-nematic case, the isotropic fluid phase and the nematic liquid crystal are both  statistically invariant under any translation in space. But for the orientational part, the two phases do not share the same symmetry group. 
Indeed, the isotropic fluid phase is statistically invariant under any rotation in three dimensions, that is,  under the elements of the group  $SO(3)$, while in the mesophase the director field plays the role of a symmetry axis, restricting the symmetry to statistical invariance under the elements of the group $SO(2)$. But for energetic reasons, the dipoles borne by the nematogens tend to align anticollinear, such that the assembly of rods is unchanged when inverting heads and tails (this dimeric structure was confirmed early by X-ray diffraction experiments in 5CB and 7CB \cite{leadbetter1975}). Hence, the full symmetry group of the nematic phase is $O(2)$.  

Many important features of a phase transition with a spontaneous symmetry-breaking are encompassed within the topology of an abstract object, called the order parameter space $\mathcal{M}$. For a phase transition with a symmetry-breaking pattern $G\rightarrow H$, the order parameter space is a manifold defined from the coset $\mathcal{M}=G/H$. The toolbox of algebraic topology (Poincaré’s former analysis situs) can then be used to seek the algebraic invariants (numbers, groups, rings…) of $\mathcal{M}$ and to classify this space into equivalence classes.

Among the many entry points, homotopy theory is of particular interest to determine the presence of singularities. Two topological spaces are homotopic if they can be mapped into each other by a continuous deformation where bijectivity is not necessarily preserved  (i.e. gluing, shrinking or fattening the space is allowed). Homotopy groups, denoted generically as $\pi_k(\mathcal{M})$, have been extensively studied in condensed matter physics, mainly in the pioneering works of Kleman, Lavrentovich, Michel, Toulouse and Volovik \cite{volovik1976line,kleman1977classification,michel1978topological,michel1980,volovik1983topological,kurik1988,kleman1989defects} and they are associated to different kinds of topological properties for $\mathcal{M}$. For instance, $\pi_1(\mathcal{M})$ tests the simple-connectedness of the order parameter space. Indeed, consider first $\mathcal{M}=\mathbb{R}\times\mathbb{R}$. It is simply-connected as all closed loops (dimension 1) are homotopic to a point (dimension 0): therefore $\pi_1(\mathcal{M})=I$ and the order parameter space is simply connected. Conversely, for $\mathcal{M}=\mathbb{R^*}\times\mathbb{R^*}$, there are two equivalence classes of closed loops: those not encircling the origin, which are homotopic to a point, and those encircling the origin which cannot be shrunk into a point. Therefore $\pi_1(\mathcal{M})\neq I$ and the order parameter space is not simply connected. The 0D-hole has thus changed the homotopy content of  $\pi_1(\mathcal{M})$. Interestingly, the dimensionality of the manifold is crucial here: a loop trying to lasso a 0D-hole can succeed in 2D but will always fail in 3D. In its most general form, the fundamental result of homotopy analysis states that in dimension \textit{n}, if the \textit{k}\textsuperscript{th} homotopy group $\pi_k(\mathcal{M})\neq I$, then holes of dimension $n-1-k$ appear: such singularities are called topological defects. Strictly speaking, a defect is topological when the singular configuration of the order parameter cannot be transformed continuously into a uniform configuration. This process depends not only on the order parameter configuration but also on the dimensionality of the  {order parameter} space (due to the possibility to ``escape in the third dimension''). 
Therefore, there is also a more flexible use for the terminology ``topological defect'', referring to the singularity associated to any non-trivial homotopy content of the order parameter manifold, whatever its topological stability is. In the remainder of this article, we will stick to that latter meaning.

\paragraph{Zoology of topological defects in nematics}  For the isotropic-nematic phase, the order parameter space is given by $\mathcal{M}=SO(3)/O(2)\equiv S^2/Z_2$: the resulting space, called the real projective plane $RP^2$, can be pictured as a 2-sphere having its antipodal points identified. The manifold corresponding to an immersion of the real projective plane in 3D space is called a Boy surface and its topology is encompassed into its first four homotopy groups: for uniaxial nematics in 3D, these are $\pi_0(RP^2)=I$ (no domain wall), $\pi_1(RP^2)=Z_2$ (existence of linear defects), $\pi_2(RP^2)=\mathbb{Z}$ (existence of point defects) and $\pi_3(RP^2)=\mathbb{Z}$ (existence of textures).   

\begin{figure}[!ht]
\begin{center}
\includegraphics[height=8.2cm]{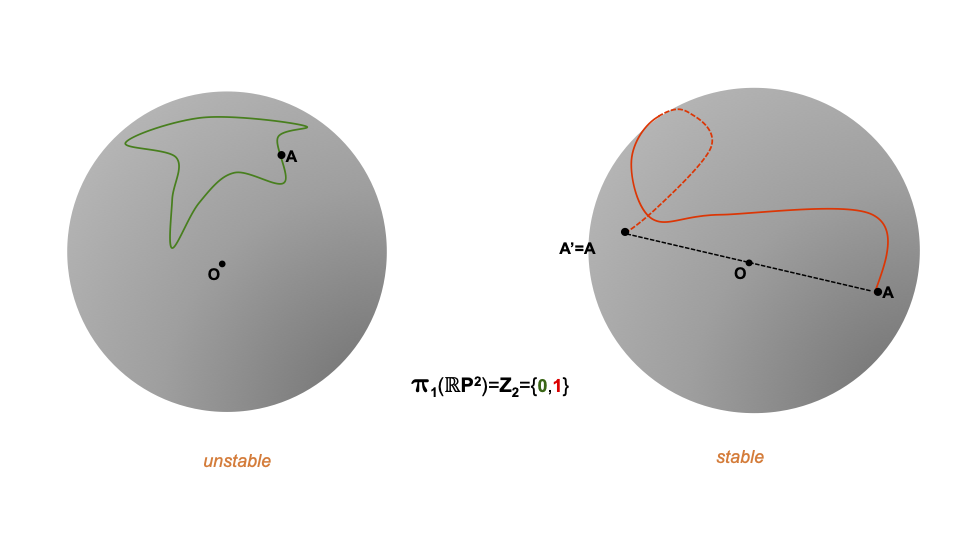} 
\caption{\textit{Left:} Class $N=0$ of closed loops homotopic to a point in the order parameter space. \textit{Right:} Class $N=1$ of closed loops consisting of paths connecting two antipodal points, which are not homotopic to a point.}\label{Z2-homotopy}
\end{center}
\end{figure}

Linear defects (or ``disclinations'' in Frank's terminology) come from a breaking of the rotational symmetry group and they are probably the most widespread singularities observed in nematics. In optical microscopy, they appear as thread-like structures used by Friedel to coin the term nematic (from the greek $\nu\eta\mu\alpha$="thread"). In polarizing microscopy, disclinations give rise to the beautiful Schlieren patterns, where dark brushes connect at singular points corresponding to the line defects viewed end on. The content of the first homotopy group (or Poincaré group) is $Z_2={0,1}$, which means that there are two equivalence classes for closed loops. The trivial class $N=0$ corresponds to defects that are not topologically stable (they can relax into a uniform configuration), whereas the second non-trivial class $N=1$ corresponds to defects that cannot be removed (see Fig. \ref{Z2-homotopy}). For reasons we will clarify later, we retain the terminology of wedge disclinations to the trivial class and the terminology of M\oe bius disclinations to the non-trivial class. A disclination can stay almost straight or form loops. It is generally associated to other disclinations within dipoles (edge dislocations), amorphous networks (blue phases), etc. In that case, they have the possibility to interconnect \cite{chuang1991cosmology} and they combine according to the algebra of $Z_2$, that is $0+0=0$, $0+1=1$ and $1+1=0$. An extensive review on linear defects in the general context of ill-ordered condensed matter can be found in \cite{kleman2008disclinations}.  Since our main concern here is disclinations we refer the reader interested in point defects and textures (including the exotic skyrmions and hopfions) to the the very complete reviews   \cite{kleman2006topological} and \cite{smalyukh2020knots}, respectively.

\subsection{Optics in the presence of linear defects}

\paragraph{Director field of axial disclinations} A region characterized by a given director field can undergo orientational distorsions as a result of external constraints. As $\mathbf{n}$ is a unit (headless) vector, the distorsions always occur in a plane orthogonal to the director field, i.e. $\delta \mathbf{n}.\mathbf{n}=0$. From a Taylor expansion, one can rewrite the deformed state as the sum of three main elastic modes: a splay term in $\boldmath{\nabla}.\mathbf{n}$, a twist term in $\mathbf{n}.(\boldmath{\nabla}\times\mathbf{n})$ and a bend term in $|\mathbf{n}\times(\boldmath{\nabla}\times\mathbf{n})|$. The Frank-Oseen free energy density is the elastic cost of orientational oscillations around $\mathbf{n}$:
\begin{equation}
    f_V=\frac{1}{2}
    K_1
    ({\boldmath\nabla}.\mathbf{n})^2
    +\frac{1}{2}K_2
    (\mathbf{n}.({\boldmath\nabla}\times\mathbf{n}))^2
    +\frac{1}{2}K_3
   (\mathbf{n}\times({\boldmath\nabla}\times\mathbf{n}))^2 . \label{FOFE}
\end{equation}
Equilibrium state corresponds to configurations such that $f_V$ is extremal. Elastic constants are of order $E_0/L$, with the interaction energy about $E_0\approx 0.1$ eV and $L\approx 1$ nm, it is customary to perform the one-constant approximation for which $K_1\approx K_2 \approx K _3=K=10^{-11}$ N. This assumption is fair for most ordinary nematics: for instance, in the case of 5CB at $298$\ \!K, one measures $K_1=6.2$ pN, $K_2=6$ pN and $K_3=8.2$ pN \cite{karat1977elasticity,bradshaw1985frank}.  

The simplest class of linear defects consists in axial disclinations and they were firsly considered by Oseen \cite{oseen1933theory} and Frank \cite{frank1958liquid} (for the class of perpendicular disclinations, proposed by de Gennes, see for instance \cite{stephen1974physics}). Orientation of the director field is ill-defined along a line (say the $z-$axis) and $\mathbf{n}$ lies in a plane orthogonal to the defect axis (in our example, the $x-y$ plane). In cylindral coordinates, let $\psi(\mathbf{r,\theta})$ be the angle between the director field and the radial unit vector. Then the Euler-Lagrange equations corresponding to a minimum of $f_V$ simply writes as $\Delta \psi=0$. The solutions representing disclination lines are given by $\psi(\theta)=m\theta+\psi_0$, where $m$ is the defect strength or topological charge (a priori in $\mathbb{R}$) and $\psi_0$ a constant phase term. Around a closed loop, the total change in $\psi$ is thus $2 m \pi$. For the director field to be well-valued, this variation is tied by the hodograph rule coming from the $Z_2$ symmetry of the nematic phase:
\begin{equation}
    \oint_{\theta=2\pi}d\psi=2 m \pi=k \pi
\end{equation}
where $k \in \mathbb{Z}$. Hence, the director field writes as \begin{equation}
    \mathbf{n}=
    \begin{pmatrix}
    \cos(m\theta+\psi_0)\\ \sin(m\theta+\psi_0)\\ 0
        \end{pmatrix}
,
\end{equation} with the topological charge constrained to be integer and half-integer, i.e. 
$m=\pm 1/2$, $\pm 1$, $\pm3/2$,\dots 

Disclinations with integer strengths are topologically removable and belong to the $N=0$ homotopy class: defects $m=+1$ and $m=-1$ are topologically equivalent and can be transformed into one another by continuous deformations. They appear in optical microscopy as thick lines and their core is not singular (possibility to escape into the third dimension). Disclinations with half-integer strengths are not topologically removable and belong to the $N=1$ homotopy class. In this latter case, fibring  over a circle about the defect line by a line segment containing the director which is met at that point, gives a M\oe bius ribbon which twists along the loop an odd number of times \cite{kleman1989defects} (on the contrary, for the $N=0$ disclination, one gets an ordinary ribbon with two sides). They appear in optical microscopy as thin lines and they display a singular core structure. As the free energy density varies in $m^2$ and therefore, it is energetically more favorable for a wedge disclination to decay into two M\oe bius disclinations, as prescribed by the combination $0=1+1$. It must be remarked that besides $\lvert m \rvert $, other topological invariants (such as the self-linking number, Poincaré-Hopf's index...) are needed to characterize the topology of a linear defect, as a disclination can globally self-connect, entangle with itself...        

\paragraph{The secrets of Fermat-Grandjean principle} 

In the geometrical optics limit, light propagates along paths that can be traveled within the least time. In the case of isotropic media, this variational formulation takes the form of the well-known Fermat's principle, established by Pierre de Fermat in 1662. In anisotropic uniaxial media, the constitutive relations involve a dielectric tensor that displays two different principal permittivities, namely $\varepsilon_{\bot}$ and $\varepsilon_{\Vert}$ (in a nematic, $\varepsilon_{\Vert}$ corresponds to the permittivity in the direction of the director field, whereas $\varepsilon_{\bot}$ is the permittivity orthogonally to it). Fresnel's equation then provides two modes inside such material: the ordinary mode, behaving similarly as in an isotropic medium with refractive index $n^2=\varepsilon_{\bot}$, and the extraordinary mode which experiences a direction-dependent refractive ray index given by \cite{landau2013electrodynamics}:
\begin{equation}
    N_e(\mathbf{r})=\sqrt{\varepsilon_{\bot}\cos^2 \beta(\mathbf{r})+\varepsilon_{\Vert}\sin^2 \beta(\mathbf{r})}
\end{equation}
where $\beta$ is the angle between $\mathbf{n}$ and the local tangent vector $\mathbf{T}$. In 1919, Grandjean extended Fermat's principle to uniaxial media and he showed that the energy carried by extraordinary light rays propagates along paths obeying \cite{oswald2005nematic}
\begin{equation}
    \delta\left(\int N_e(\mathbf{r}) d\ell\right)=0
\end{equation}
where $\ell$ is the curvilinear abscissa that parameterizes a ray. 

\begin{figure}
\begin{center}
\includegraphics[width=17.5cm]{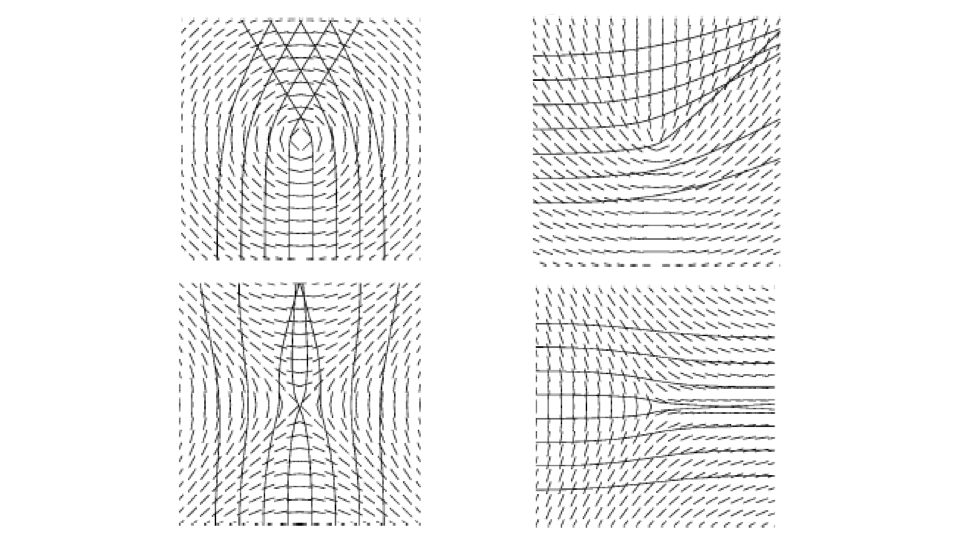} 
\caption{Light paths and director fields in the presence of a planar disclination (Up left: $m=1, \psi_0=\pi/2$. Down left: $m=-1, \psi_0=\pi/2$. Up right: $m=1/2, \psi_0=\pi/4$. Down right: $m=-1/2, \psi_0=0$). Taken from \cite{satiro2006lensing}.}\label{Fermat-Grandjean}
\end{center}
\end{figure}

Because the director field changes from point to point, a nematic generally displays a varying refractive index and hence, extraordinary light beams propagate into the medium along curves (see Fig. \ref{Fermat-Grandjean}). In the case of planar axial disclinations, the direction of $\mathbf{n}$ and consequently $\beta$ is known at each point. In that case, it can be shown that the integrand in Fermat-Grandjean's principle can be generally rewritten as \cite{satiro2006lensing,satiro2008deflection} 
\begin{eqnarray}
    N_e^2(\mathbf{r}) d\ell^2&=&\left(\varepsilon_{\bot}\cos^2\left[ (m-1)\theta+\psi_0\right]+\varepsilon_{\Vert}\sin^2 \left[ (m-1)\theta+\psi_0\right]\right)dr^2 \nonumber \\
    &&+\left(\varepsilon_{\bot}\sin^2\left[ (m-1)\theta+\psi_0\right]+\varepsilon_{\Vert}\cos^2 \left[ (m-1)\theta+\psi_0\right]\right)r^2d\theta^2 \nonumber \\
    &&-\left(\varepsilon_{\Vert}-\varepsilon_{\bot}\right)\sin^2\left[ 2(m-1)\theta+2\psi_0\right]r dr d\theta+dz^2 \label{general-discli}
\end{eqnarray}

In a seminal work \cite{gordon1923lichtfortpflanzung}, Walter Gordon pointed out the formal analogy between light propagation inside a moving dielectric and light propagation inside a non-Euclidean geometry. This idea was developed by many authors eversince \cite{quan1957inductions,plebanski1960electromagnetic,evans1986f,nandi1995optical,evans1996optical,alsing1998optical,leonhardt2006general}, as it elegantly replaces the resolution of Fermat's principle in a material medium by the search for the minimum-length lines (or geodesics) of an empty curved space. The main asset of that point of view is that one can use the toolbox of differential geometry to understand how the defect modifies transport phenomena in its vicinity. To illustrate how this works, let us consider the example of a ($m=1, \psi_0=\pi /2$)-disclination (see Fig. \ref{Fermat-Grandjean}). For this defect, Eq. \eqref{general-discli} leads to the following line element:
\begin{eqnarray}
ds_{3d}^2=N_e^2(\mathbf{r}) d\ell^2=dr^2+\alpha^2r^2d\theta^2+dz^2 , \label{wedge-discli}
\end{eqnarray}
where $\alpha^2=\varepsilon_{\Vert}/\varepsilon_{\bot}$.
The line element is a fundamental quantity in differential geometry and it simply consists in a generalization of Pythagoras' theorem for computing distances in arbitrary geometries. Here, instead of the familiar Euclidean line element $ds_{3d}^2=dr^2+r^2d\theta^2+dz^2$, the term in $\alpha^2$ means that the circumference of a closed unit circle about the defect is no longer $2\pi$ but $2\pi\alpha$ instead: in other words, there is a mismatch angle (called Frank angle) of value $2\pi(1-\alpha)$ compared to flat geometries. It is customary in differential geometry to rewrite the line element as $ds_{3d}^2=g_{ij}dx^i dx^j$, where Einstein's summation convention on repeated indices is used. $g$ is called the metric tensor and it corresponds to a positive definite quadratic form. The curvature scalar \cite{Nakahara2003} as computed from the metric is:
\begin{equation}
R=\frac{2\pi(1-\alpha)}{\alpha r} \delta^2\left(r\right) \label{curvature-discli}
\end{equation}
whereas the torsion tensor is identically zero.

An alternate approach to describe the influence of defects in optics, also from differential geometry, consists in using the formalism introduced by Paul Finsler in his 1918 thesis, for which there is no quadratic constraint on the geometry as in the Riemannian case \cite{chern1996finsler}. As a matter of fact, the arc length is given by a Finsler function $F$ such that $ds_{3d}=F\left(x,y,z,dx,dy,dz\right)$ instead of $ds_{3d}=\sqrt{g_{ij}dx^idx^j}$. In the case of anisotropic media, $F(\mathbf{r})=N_e(\mathbf{r})d\ell$  and the metric corresponds to the Hessian of the ray index \cite{joets1994geometrical}. Ought to the particularly simple dependency of the ray index with respect to coordinates, this formalism turns out to be fully equivalent to Riemann's approach (see discussion at the end of \cite{satiro2006lensing}).

\begin{figure}
\begin{center}
\includegraphics[width=16.5cm]{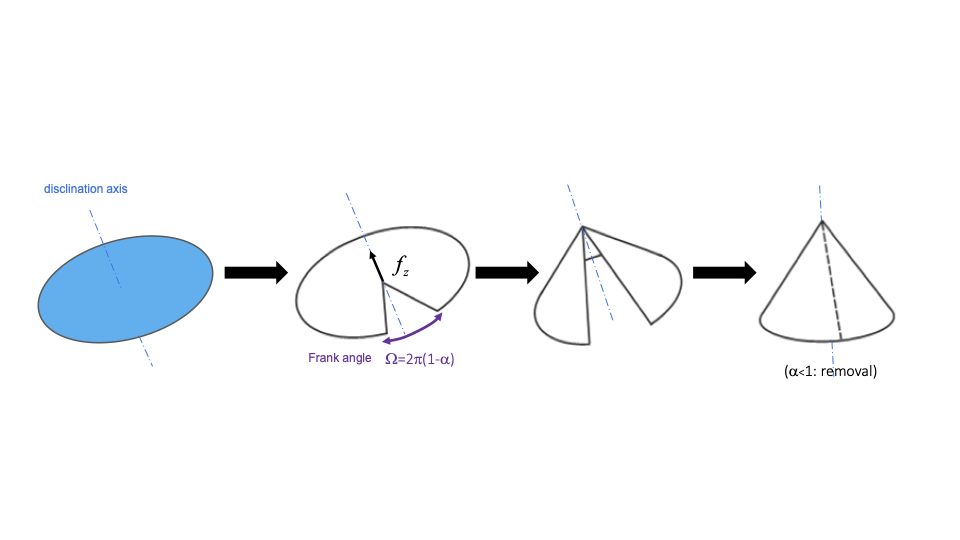} 
\caption{Volterra cut-and-weld process for a (wedge) disclination along z.}\label{volterra2}
\end{center}
\end{figure}

 A line element of exactly the same form as \eqref{wedge-discli} appears in the geometric theory of defects in elastic media \cite{katanaev2005geometric}, related to the strain field associated to wedge disclinations as will be described in Section \ref{pitfalls}. Such line defects can be formed by either inserting or removing a wedge of material of angle $2\pi(1-\alpha)$ with subsequent identification of the edges. In the case of a removal wedge disclination of axis $z$ ($\alpha<1$), the geometry surrounding the defect is conical and can be easily pictured from the Volterra cut-and-weld process of Fig. \ref{volterra2}. In other words, a disclination can be pictured as a Riemann manifold, for which curvature is only located on the disclination axis and vanishes everywhere else. 

\medskip

\paragraph{Discussion} From the elasticity point of view (as opposed to optics) the description of an axial wedge disclination by the geometry (\ref{wedge-discli}), or more generally by \eqref{general-discli}, calls for several remarks (it is important to stress here that $\varepsilon_{\Vert}$ and $\varepsilon_{\bot}$ now are related to elastic, not optical, anisotropy). First, a liquid crystal consists in an assembly of rod-like molecules and modeling it as a continuous medium is not self-explanatory. Rigorously, the continuum limit for nematoelasticity should come as a coarse grained approximation of molecular dynamics and it should fail at the atomic scale. $\boldsymbol{n}(\mathbf{r})$ is defined statistically, as the average common direction of the nematogens at each ``point'' in space. The ``point'' actually refers to a small volume of space that includes enough molecules for the averaging process to be physically significant. Hence, in practice, it means that the ``point-volume'' has to be large enough compared to the molecular scale $a$ (typically $a\approx 20$ \AA) and that the variations of the director field must occur at much larger scales than $a$. Only then, the distorted liquid crystal can be described as a continuous medium, as discussed by Oseen \cite{oseen1933theory}, Zöcher \cite{zocher1933effect} and Frank \cite{frank1958liquid}.  

A second caveat is related to the status of  (\ref{wedge-discli}), which obviously possesses non-vanishing curvature as in three-dimensional gravity. Yet, the nematic actually lives in a three-dimensional Euclidean space, which means that the background geometry is flat. How to reconcile these two standpoints? Following the analysis from De Wit \cite{de1981view}, the state described by (\ref{wedge-discli}) does exist in the flat space, but only in an imaginary space where the medium is relaxed: $g_{ij}$ comes from the projection of this imaginary space onto the physical flat space, in a similar way as a stereographic map projection transfers the geometric properties on a 2-sphere (the Earth, with its meridians and parallels) onto a flat plane while deforming them (Wulff net). It turns out that the geometric description of defects thus requires two metrics: 1) The physical flat metric, $\delta_{ij}$, will be used to perform operations on tensors such as raising/lowering indices...  2) The effective metric $g_{ij}$, which contains the elastic information, will be used to determine the kinematics of low energy perturbations (geodesics, first integrals...). 

Third, one may naturally wonder what really happens on the defect axis and how to refine our zero-width model. In soft matter and more especially nematics, defect cores are very narrow as well but they still belong to the realm of continuum mechanics. As discussed in \cite{oswald2005nematic}, a disclination line can accurately be described by a ``two-phase model'': the core consists in a tubular region, filled with the nematogens in isotropic phase (vanishing order-parameter), and surrounded by the nematic phase (non-zero order parameter). This approach is consistent with exact solutions obtained from the minimization of the Landau -- Ginzburg -- de Gennes free energy. Yet, the last word has probably not been said about disclination cores: observations made on lyotropic chromonic liquid crystals revealed that the core region has several unexpected features (asymmetric non-circular interfaces between the nematic and the isotropic phases, azimuthal and radial dependencies for the phase and amplitude of the order parameter...) compared to classic two-phase models \cite{zhou2017fine}.

\subsection{Analogue gravity: lessons and pitfalls}
\label{pitfalls}

\paragraph{Physics as geometry} The geometric description of transport near linear defects does not restrict to optics near axial disclinations. Since the pioneering works by Bilby \cite{bilby1955continuous} and Kröner \cite{kroner1958kontinuumstheorie} in the 1950s, this approach has been extended to elasticity theory as well \cite{krivoglaz1983physics,kadic1983gauge}. In the noteworthy set of works \cite{katanaev1992theory,katanaev1999scattering,katanaev2005geometric}, Katanaev proposed a general framework based on Riemann-Cartan manifolds for dislocations and disclinations in elastic media but only considered the strain tensor field as the relevant degree of freedom. An expression of the effective metric $g_{ij}$ in the medium rest frame can be obtained in the case of linear elasticity as
\begin{equation}
g_{ij}=\delta_{ij}+2\varepsilon_{ij} \label{analog-elasticity}
\end{equation}
where $\varepsilon_{ij}$ denotes the strain tensor. Compared to ordinary elasticity theory (OET), the geometric theory of defects is in principle more accurate (ordinary elasticity only reproduces the first-order approximation of the geometric theory of defects \cite{katanaev2005geometric}) and it is more versatile (changing the kind of defect only requires changing the metric, instead of a complicated set of boundary conditions in ordinary elasticity theory). Moreover, the geometric approach is also likely to encompass many other kinds of linear defects of interest in liquid crystals physics, such as screw dislocations in smectic A and C \cite{pershan1974dislocation,kleman1980screw} (in that case, the defect must be described in terms of a Riemann-Cartan manifold, for which torsion is only located on the dislocation axis and vanishes everywhere else \cite{de1998geodesics}), dispirations in antiferroelectric Sm$C_A$ and the dimeric Sm$C_2$ \cite{takanishi1992visual,de1998geodesics}, edge dislocations in smectics \cite{meyer1978observation,moraes1996geodesics} (which are merely disclination dipoles)...

The preceding examples testify that in many condensed matter systems, the effective degrees of freedom are represented by specific field excitations that propagate over effective Riemann-Cartan manifolds. Geometrization of physics is not a new idea. In Plato's \textit{Timaeus}, an attempt was made to describe the world in terms of only five regular polyhedra and ever since, geometrization of physics has been a dream pursued by many figures in science, including René Descartes, Bernhard Riemann, William K. Clifford \cite{clifford1976space} (for an updated account, see \cite{elizalde2013bernhard})... The most successful step forward in merging geometry and physics was made in the twentieth century by Albert Einstein with the theory of general relativity: the gravitational interaction turns out to be nothing more than a manifestation of the spacetime curvature. The possible implications of that theory did not escape the attention of influencial physicists such as Hermann Weyl \cite{weyl1919neue}, Arthur Stanley Eddington \cite{eddington1922space} and more especially John A. Wheeler. In the seminal paper \textit{Classical physics as geometry} \cite{misner1957classical}, Wheeler and Charles W. Misner borrow tools from cohomology, differential geometry, exterior algebra and topology to fully merge gravitation, electrodynamics and geometry. Provided spacetime is multiply-connected, Misner and Wheeler showed that similarly to mass, classical charge can also be seen as a byproduct of the spacetime geometry. A particularly meaningful example is the low-dimensional gravitational model proposed by Gerard 't Hooft in the context of quantum gravity \cite{t2008locally} (see below \ref{BHEU}): in 2+1 dimensions, 't Hooft showed that gravitating point particles can elegantly be described as  conical point-like singularities of space-time, each deficit angle being related to the particle's total energy. Today, this kind of ideas has spread out to the point where it has become an area of research on its own: analogue gravity (for an extensive review, see \cite{barcelo2011analogue} and more recently \cite{jacquet2020next}).

\paragraph{Pitfalls} Despite appealing to classical fields, analogue gravity is tricky and must be handled with great care. Indeed, it relies simultaneously on two different manifolds: 1) the background gravitational metric -- which is experienced by all fields (generally Minskowski's) -- is the outcome of Einstein-Cartan equations. It is a tensor, used for instance to raise and lower indices of tensors, and as such, it is a covariant quantity, and 2) the effective metric -- which is experienced only by the fields coupled to matter -- does not obey Einstein-Cartan gravitational equations. Its purpose is limited (for instance, to determine the geodesics followed by the coupled-fields excitations, as it is not a covariant quantity. Indeed, the effective metric is derived from physical quantities which are defined in a privileged frame, the medium rest-frame, and that are not invariant under Lorentz transformations.

As can be seen from (\ref{analog-elasticity}), the effective metric superimposes the background Minkowski metric and a correction taking into account the couplings between field and matter. In the original experiment led by Hippolyte Fizeau, the changes in the velocity field of water (and hence of the Gordon metric itself) were obviously ruled by the Navier-Stokes equations (for the velocity field) instead of Einstein-Cartan equations. In other words, effective spacetimes are generally stationary.  Ref~\cite{simoes2010liquid} pointed out that textures in nematic liquid crystals can indeed be described by the space sector of an Einstein-like equation, with the elastic-stress tensor replacing the energy-momentum tensor. The relevance of the effective metric is therefore restricted to calculations of properties related to the \textit{kinematic properties} of the fields coupled to matter. This encompasses as we said the geodesics of low-energy excitations but also the less obvious cases of Unruh effect or Hawking radiation which are purely kinematic phenomena \cite{visser2003essential}. Therefore, the analogy between gravitation and condensed matter is strictly kinematic but not dynamical. To rephrase Wheeler, \textit{analog spacetime tells matter how to move... but matter does not tell analog spacetime how to curve}.

What is the purpose of analogue gravity? In cosmology, putting a theory into test is always a thorny challenge. In 1992,  the great epistemologist Karl Popper already pointed out that the ``\textit{major theoretical problem in cosmology is how the theory of gravitation may be further tested and how unified field theories may be further investigated}'' \cite{popper1994}. If the plentiful harvest of low-energy observations (baryonic oscillation spectroscopy, gravitational wave interferometry, mapping of the cosmic background ...) answered many questions, theoretical models involving (trans)planckian scales bloomed even faster, for which experimental confirmations seem almost impossible -- even in principle -- to reach. A possible way out of this conundrum is to take advantage of the richness and flexibility of condensed matter. Within certain limits, analogues of gravity can be used to simulate different types of cosmological objects (signature transitions events, cosmic strings...) and to investigate the transport of bosonic and fermionic quasiparticles in nontrivial spacetimes. The next section reviews a series of works dealing with non-standard cosmological models that can be investigated from their liquid crystal counterparts.

\section{Unraveling the Universe with liquid crystals: cosmology in the laboratory}

\subsection{Phase transitions in cosmology}

\paragraph{Thermal history of the universe} In many senses, cosmology  consists in thermodynamics applied to the largest expanding closed system: our universe. Our current understanding of cosmic history is indeed based on the Standard Hot Big Bang Model and it originates in the pioneering works of three founding fathers: Albert Einstein, Alexander Friedman and George Lemaître. In essence, this model states that about 13.8 billion-years ago, the Universe was in an extremely hot dense state, consisting in a quark-gluon plasma, and that it has expanded ever since. In the framework of grand unified theory (GUT), the four fundamental interactions (the gravitational interaction, the electromagnetic interaction and two lesser known forces, the weak nuclear interaction -- responsible for radioactive $\beta$ decays -- and the strong nuclear interaction -- which ensures the cohesion of the atomic nuclei) were then assumed to be unified at energy scales estimated at about $10^{16}$ GeV. 

Each interaction is associated with internal (or gauge) symmetries: for instance, at today's energy scales, the electromagnetic force displays gauge invariance under the elements of $U(1)$, the unitary group of dimension 1 (for an accessible review on gauge theories see for instance \cite{jackson2001historical}). Above $10^{16}$ GeV, the group \textit{G} containing the internal symmetries of grand unified superforce is not known for sure and many candidates with exotic names are considered, such $SU(5), SU(6), SU(7), SU(8), SU(9), SO(10), SO(14), E_6$... \cite{peter2009primordial} The universe expansion played the role of a gigantic Joule-Thomson expansion, which caused a large temperature drop driving cosmological phase transitions. For example, the last of these transitions is the electroweak phase transition, occurring at energy scales about $10^2$ GeV. It marks the splitting of the electroweak force into an electromagnetic part, described by Maxwell's theory (1865), and the weak nuclear part, the first theory of which being Fermi's theory (1933). This transition involves a spontaneous gauge symmetry breaking: the high temperature gauge symmetry group $SU(3)_c\times SU(2)_L \times U(1)_Y$ broke into $SU(3)_c\times U(1)_{em}$ \cite{peter2009primordial}.

Let us now examine the topology of vacuum manifold (that is the set of field configurations minimizing the free energy modulo gauge transformations), which is the equivalent of the order parameter space $\mathcal{M}$ in condensed matter physics. In \cite{jeannerot2003generic}, Jeannerot et al determined the homotopy content corresponding to all eligible groups \textit{G} likely to decay below $10^{16} $GeV into $SU(3)_c\times SU(2)_L \times U(1)_Y$. Their conclusion leaves no doubt concerning the formation of cosmic strings: 

\begin{quotation}\textit{\dots among the SSB schemes which are compatible with high energy physics and cosmology, we did not find any without strings after inflation.}
\end{quotation}
(if one assumes that the universe is topologically multi-connected, cosmic strings and monopoles may appear -- not single but pairwise --, whereas two-dimensional defects -- domain walls -- cannot form at all \cite{uzan1997no}). 

\paragraph{Kibble-Zurek mechanism} Cosmic inflation is a period of extremely fast expansion of the Universe scale factor (typically a factor $10^{26}$ within $10^{-32}$ seconds) that presumably happened at the very beginning of the universe \cite{guth1981inflationary}. From the point of view of statistical physics, inflation is nothing more than a quench and as such, it is likely to favor the formation of topological defects. In 1976 \cite{kibble1976topology}, Tom Kibble introduced a three-step mechanism (later refined by Zurek \cite{zurek1996cosmological} who included the sensitivity to the quench speed) to describe the details of this quench. Basically, the Kibble-Zurek mechanism (KZM) consists in a nucleation process very similar to what happens at the isotropic-nematic phase transition, but instead of having an order locally described by the director field $\mathbf{n}$, it is here described by the phase of a complex scalar field generically called a Higgs field -- or an inflaton, because it needs not be the Higgs field responsible for the later breaking of electroweak symmetry. First, ordered protodomains (analog to the nematic spherulites) with no correlation between each other are formed and at the scale of a whole protodomain, the fast temperature drop due to inflation causes the Higgs field to locally take a non-vanishing vacuum expectation value and hence to make a phase choice. Then the protodomains grow in size until they coalesce. But as they were not correlated, the choices for the Higgs phase (technically, its vacuum expectation value) do not match in general, and line singularities of the Higgs appear when the boundaries of protodomains finally meet. These linear singularities are called cosmic strings.

Besides this qualitative predictions, the KZM also makes quantitative predictions such as the scaling dynamics of the cosmic string network, the average density of defects, correlations between defects and antidefects... In the 1990s, several works \cite{chuang1991cosmology,bowick1994cosmological,digal1999observing,kibble2007phase,mukai2007defect,repnik2013symmetry} showed that the KZM, originally developped for cosmology, was also perfectly describing line defects in nematics with the very same scaling coefficients. For instance, this model predicts that \textcolor{black}{in 2D} the density of strings scales as $\rho \sim \left(t/\tau_q\right)^{\alpha}$ with a critical exponent $\alpha_{th}=0.5$, and measurements done by \cite{chuang1991cosmology} with 5CB indeed gave $\alpha_{th}=0.51\pm0.04$. To sum up: defects consist in regions that cannot relax into the new vacuum or equivalently that are unable to make the transition into the new ordered phase, and they occur during phase transitions in cosmology and in liquid crystal physics that seem to belong to the same universality class. But the family resemblance goes further. Networks of cosmic strings and networks of disclinations also share similar intersection processes: 1) when two line defects intertwine, they may reconnect the other way as they cross (intercommutation) \cite{vilenkin1985cosmic,chuang1991cosmology} and 2) when one line defect self-intersects, it creates a loop \cite{brandenberger1994topological,duclos2020topological}.

\paragraph{An almost perfect analogy?} Last but not least of these common points: the geometry. Nambu-Goto strings, which are the simplest cosmic defects one may expect in cosmology, consist in linear concentrations of energy and as such, they are considered as infinitely thin objects (as the thickness of a cosmic string is estimated at $10^{-28}$ cm, this is a fair approximation\footnote{Compare, for instance, with the Bohr radius $a_0=5.29 \times 10^{-9}$ cm.}). As required by thermal field theory and general relativity, the geometry around a Nambu-Goto string is described by the Vilenkin's line element \cite{vilenkin1985cosmic}: 
\begin{equation}
    ds^2=-dt^2+dr^2+\left(1-4 G\mu\right)^2r^2d\theta^2+dz^2 \label{Vilenkin}
\end{equation}
where $\mu$ is the string energy density estimated at about 10 million billion tons per meter (we adopt hereafter the customary unit system of cosmology where $c=1$). The space part of this element is identical to (\ref{wedge-discli}): it is a conical geometry corresponding to a removed Frank angle \footnote{However, no isolated cosmic string corresponds to the case of an added Frank angle (saddle-like geometry). To our knowledge, the existence of a negative-mass cosmic string has been considered only when the defect is associated to a wormhole \cite{visser1989traversable, cramer1995natural}.} (typically, for a GUT scale string, this angle is a few seconds of arc).
The reader interested in the classical gauge theory of string interactions in curved spacetimes can refer to Ref.\cite{PhysRevD.105.105026}. From the standpoint of the soft-matter physicist, Nambu-Goto strings can be understood as the cosmic counterparts of wedge disclinations. How to make sense of such incredible similarity? For the most part, this question is still open, but a noticeable attempt to address it was done in \cite{simoes2010liquid}: in essence, the reason is that equations of nematoelasticity have the form as the spatial sector of Einstein’s field equations, with the elastic-stress tensor playing the role of the energy momentum tensor. 

As the analogy between gravity and nematoelasticity does not concern time components, one expects that the dynamics of a cosmic defect cannot be directly mapped with those of a disclination. There are other discrepancies between cosmic and elastic defects that one must bear in mind to avoid fallacies. Obviously, the motion of disclinations is classical (typically a few $\mu$m per second) whereas cosmic strings are ultra-relativistic. Dissipation mechanisms for cosmic strings are due to radiation of gravitational waves, while those in liquid crystals are friction-dominated. What is the outcome on the dynamics of the defects? In cosmology, monopoles annihilate in pairs (Langacker-Pi mechanism), but they do not annihilate fast and early enough to avoid that the Universe becomes monopole dominated (which is why inflation is necessary, as it drives monopoles very far away from each other).  On the contrary, elastic hedgehogs in a nematic annihilate rapidly according to a scaling law. At a more fundamental level, this is linked to the fact that in high energy physics, broken symmetries are gauged (or internal) whereas in liquid crystals, broken symmetries are geometrical: in the first case, one is dealing with ``gauged defects'' and in the second case, one is dealing with ``global defects''. 

\subsection{Beyond cosmic wedge disclinations}\label{BWD}

\paragraph{The way out of an observational dead-end} Cosmic wedge disclinations exist either as stable infinite straight lines (their equation of state simply equates the string energy density to its tension $\mu=T$) or as closed loops that radiate away gravitational waves until they vanish. When moving, strings happen to distort spacetime such that at all scales, matter accretes along its wake into sheet-like structures. They may account for the formation of large-scale structures in our universe (including the Great Wall) and they have several expected observable signatures such as the Kaiser-Stebbins effect \cite{kaiser1984microwave,vilenkin1986looking} (an asymmetric Doppler shift giving rise to anisotropies of the cosmic microwave background), gravitational lensing \cite{Vilenkin1994} (not in the form of an Einstein ring, but as a double image instead), geometric phase (Aharonov-Bohm effect but with a cosmic string replacing the flux tube \cite{bezerra1987gravitational})... Up to now, data collected by the PLANCK mission (2014) only settle upper bounds on the string parameter $\mu$ \cite{ade2014planck} and in 2020, observations of the stochastic gravitational wave background (NANOGrav experiment) may have provided with first evidences for cosmic strings \cite{buchmuller2020nanograv,blasi2021has,ellis2021cosmic}. 
   
The non-conclusive observations of Nambu-Goto strings call for the search of refined models for linear defects. In fact, the zero-width approximation and the straightness of cosmic strings are probably too coarse to account for realistic defects. Hiscock \cite{hiscock1985exact} and independently Gott \cite{gott1985gravitational} suggest to smoothen this singularity by introducing two string models with a core structure of constant curvature: the flower-pot model (with zero curvature) and the ballpoint-pen model (with non-vanishing curvature). In the Gott-Hiscock thick cosmic string spacetime, the metric tensor is piecewise-defined and it must obey matchings conditions at the core radius \cite{allen1990effects}: the extrinsic curvature of the boundary should be the same whether measured in the interior or exterior region (O'Brien-Synge-Lichnerowicz jump condition). In contrast, thanks to experiments \cite{zhang2005nature,zhou2017fine} and  molecular simulations \cite{andrienko2000molecular,harkai2021manipulation}, much is known about the NLC disclination core. In particular, there is strong evidence for biaxiality and that strength $+1$ disclinations are in fact bound pairs of strength $+1/2$ ones, which may be manipulated by electrical fields \cite{susser2020transition}. We note that this rich structure may serve as an inspiration for novel cosmic string core models.  In the same line, instead of being perfectly straight, linear defects can present cusps, kinks and wiggles: the averaged effect of these perturbations increases the linear mass density $\mu$ and decreases the string tension $T$, as prescribed by the equation of state $\mu\:T=\mu_0^2$ \cite{peter1994comments,carter1995transonic,Frank2017wiggly}. Compared to straight string, the geometry remains conical but the deficit angle is larger than in the straight string case, which increases polarization anisotropies of the cosmic background radiation \cite{pogosian1999cosmic}. 

There are many other ways to dress a Nambu-Goto string such that it may account for observational results \footnote{We will consider here only models likely to have a liquid crystal analog, which -to our knowledge- should exclude superconducting cosmic strings.}. From an extension of Volterra process to 3+1 dimensions, Puntingam and Soleng showed that there was only 10 ways to modify a Minkowski spacetime into different pseudo-Riemann–Cartan geometries with respect to the Poincaré group. For example, a cosmic linear defect can display chirality \cite{gal1993spinning,dias2005effects,wang2015deformations,vitoria2018rotating,zare2020duffin,ahmed2020quantum}: in that case, the defect carries torsion along its axis and one gets the cosmic counterpart of a screw dislocation in a smectic liquid crystal. Twisted Nambu-Goto strings (or cosmic dispirations), consisting in spacetimes with delta function-valued curvature and torsion distributions have also been considered, as they combine both rotational and translational anholonomy \cite{de2003renormalized,cai2018radiative,mota2018scalar,de2021scalar}: as mentioned earlier, their effects on light could be tested from experiments done with elastic dispirations Sm$C_A$ and Sm$C_2$.

\paragraph{Going further}
Rotating disclinations are not likely to be stable but it is worth mentioning here that their cosmic counterparts have been long predicted in the literature \cite{xanthopoulos1986rotating}. A metamaterial analogue of the rotating cosmic string spacetime has  been proposed \cite{mackay2010towards}, as well as  a superfluid vortex analogy \cite{volovik2001superfluid}. One of the most interesting properties of the spinning string is its association to closed timelike curves which may find applications in time travel \cite{jensen1992notes}. Incidentally,  parallel cosmic strings moving in opposite directions have been suggested \cite{gott1991closed} as a prototype time machine. Related to, but not really a model for spinning strings, is the case of  a  hyperbolic nematic-based metamaterial with a disclination which was addressed in \cite{fumeron2015optics}. Along the same line, in \cite{figueiredo2017cosmology} it was proposed a disclination model for the compactified Milne model of a cyclic universe. More details on this model in Section \ref{BHEU}.

 Among the gauge strings, a very interesting possibility is the semilocal string \cite{hindmarsh1993semilocal}. Like Dirac's string of magnetic dipoles, semilocal strings end on gauge monopoles. They are analogous to real disclinations in liquid crystals which, due to the finite size of a liquid crystalline sample, must end somewhere (hedgehog, 2D disclination on the liquid surface or on receptacle wall) or else, form a loop. 
Apropos, disclination loops in active nematics have very complex dynamics (including chaos) and may present recombination episodes \cite{duclos2020topological}.

 \textcolor{black}{Defects in liquid crystals have inspired many other proposals in cosmology. GUT allows for discrete gauge symmetry groups, the standard $Z_2$ parity and one $Z_3$ parity, which are the only anomaly free groups that remain unbroken at low energy \cite{ibanez1992discrete,Vilenkin1994}. The corresponding cosmic strings are generically called $Z_n$-cosmic strings. Based on the known physics of Moebius disclinations, which are commonly observed in nematics, Satiro and Moraes have investigated some cosmological outcomes of $Z_2$ cosmic strings \cite{satiro2009asymmetric}: in particular, they showed that $Z_2$-cosmic strings display both positive and negative mass density regions. Bearing in mind the back and forth interplay between cosmology and soft matter, one cannot avoid to mention the latest works of Maurice Kleman, who imported homotopy theory from condensed matter to astrophysics and cosmology \cite{kleman2009forms,kleman2014tubes,kleman2015some}. In particular, he conjectured and classified new families of cosmic defects (such as r-cosmic forms) allowed in a four-dimensional maximally symmetric spacetime \cite{kleman2015some}.}

\subsection{Black holes and Early universe}\label{BHEU}

\paragraph{Black holes, white holes, wormholes}

This is sometimes referred to as the ``cosmology in the laboratory'' game plan and it covers topics such as classical black holes \cite{Garay2001}-\cite{Wilczek2020}, Hawking radiation \cite{Unruh1981}-\cite{Nova2019}, wormholes \cite{Part2015}-\cite{azevedo2021optical}... For instance, in Haller’s approximation, the hydrodynamics of a nematic liquid crystal radially flowing down a drainhole is experienced by light beams as the equatorial section of the Schwarzschild’s metric  
\begin{equation}
    ds^2= - \left( 1-\frac{2M}{r} \right) dt^2 + \frac{dr^2}{\left( 1-\frac{2M}{r} \right)} + r^{2}(d\theta^{2} + \sin^{2} \theta d \phi^{2}) 
\end{equation}
for a specific velocity profile. The ordinary and extraordinary indexes of the NLC depend on the scalar order parameter of
the liquid crystal. So, it was possible to taylor those indexes to get the proper optical metric.
In order to achieve this, the Beris-Edwards hydrodynamic theory wass used to connect the order parameter with the
velocity of the liquid crystal flow at each point. This was done in Ref.  \cite{pereira2011flowing}.

More recently, an optical analogue of a wormhole threaded by a cosmic string was described in \cite{azevedo2021optical}. Wormholes are solutions of Einstein's equations that connect different regions of the  spacetime. For instance, a spherically symmetric wormhole can be obtained by joining two Schwartzschild black hole spacetimes by a spherical hole carved around each singularity.   Wormholes  are usually represented by ``embedding diagrams'', which are 2D slices of the 4D structure immersed in Euclidean 3D space. The embedding diagram of the notorious Morris-Thorne \cite{morris1988wormholes} wormhole is obtained by taking a $t = \text{const.}$, $\theta = \pi/2$ section of the spherically symmetric spacetime described by the metric
\begin{equation}
    ds^{2} = - c^{2} dt^{2} + \frac{dr^{2}}{1 - b_{0}^{2}/r^{2}} + r^{2}(d\theta^{2} + \sin^{2} \theta d \phi^{2}).
    \label{MTmetric}
\end{equation}
 The restricted metric, $ds^{2} = \frac{dr^{2}}{1 - b_{0}^{2}/r^{2}}+r^{2} d\phi^{2}$, can be embedded in a 3D Euclidean space with metric $ds^{2} = dz^{2} + dr^{2} + r^{2} d\phi^{2}$ such that $z = z(r)$ is the equation of the embedded surface of revolution. For metric \eqref{MTmetric} the result is the catenoid.

\begin{figure}[htp]
\centering
    \includegraphics[width=0.29\columnwidth]{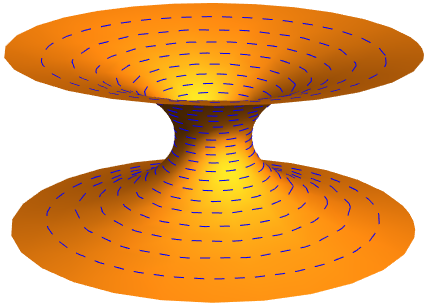}
      \includegraphics[width=0.29\columnwidth]{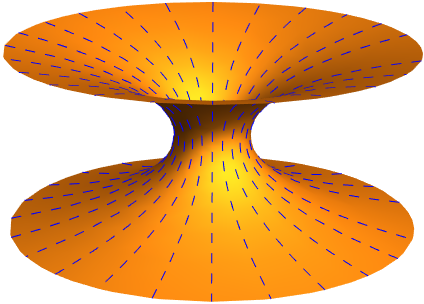}
     \caption{Director field for circular and radial $+ 1$ disclinations on the catenoid, respectively. Taken from \cite{azevedo2021optical}.}
    \label{catenoiddisclination}
\end{figure}

A thin nematic film on a catenoid  with director field aligned either circularly or radially (see Fig. \ref{catenoiddisclination}) has an optical metric given by \cite{azevedo2021optical}
\begin{equation}
    ds^{2} = d\tau^{2} + \alpha^{2} (\tau^{2} + b_{0}^{2}) d\phi^{2},
    \label{metric1.1}
\end{equation}
where $\alpha = {n_{o}}/{n_{e}}$ for the circular case, and $\alpha = {n_{e}}/{n_{o}}$ for the radial one. The coordinate $\tau$ is the arc length of the catenary that under rotation gives rise to the surface. The parameter $b_0$
is the radius of the wormhole ``throat''. For $\alpha=1$, Eq. \eqref{metric1.1} reduces to the catenoid metric. It is clear from Eq. \eqref{metric1.1} that, asymptotically ($\tau >> b_0$), the optical metric of the disclination is recovered. This is also evident from the top view of the catenoids of Fig. \ref{catenoiddisclination}.

\begin{figure*}[htp]
	\centering
	(a)\includegraphics[width=0.25\columnwidth]{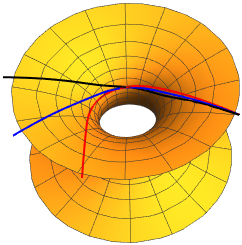}
	\hfill (b) \includegraphics[width=0.25\columnwidth]{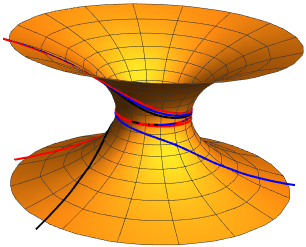}
	 \hfill (c)\includegraphics[width=0.25\columnwidth]{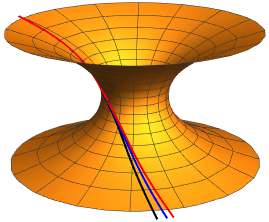}
	\caption{Assorted geodesics for the circularly decorated catenoid.  The blue  lines represent the isotropic case $\alpha = 1$. The red  and black lines represent, respectively, circular (deficit angle) and radial (surplus angle) disclinations with $\alpha = 0.85$ for (a) and (b), and with $\alpha = 0.98$ for (c). Taken from \cite{azevedo2021optical}.}
	\label{orbs}
\end{figure*}

This optical model simulates the conical spacetime of a Morris-Thorne wormhole threaded by a cosmic string. The geodesics as obtained in \cite{azevedo2021optical} are represented in Fig. \ref{orbs}.

From Fig. \ref{orbs} it is clear that the two parts of the wormhole joined by its throat act as a black hole/white hole pair.

\paragraph{Road to quantum gravity}

A major contemporary challenge in physics is to find an extension of General Relativity able to describe gravity at all energy scales, in particular at the very beginning of the universe. This is the mission devoted to quantum gravity theories, which have the daunting task of reconciling Einstein's general relativity and quantum field theory. Despite promising attempts including superstring theories, M-theory or quantum loop gravity, no proposal is entirely satisfactory up to now, and even so, the energy scales required to test these theories are far beyond our current scientific capabilities. A way out of this gridlock is to rely on simpler models that capture the essential features of quantum gravity but remain connected to low-energy-physics systems, i.e. analogue gravity. The rare pearl was first introduced in a seminal paper by Deser, Jackiw and 't Hooft \cite{deser1984three}: 2+1 gravity with point-particle sources. 

The main point is that there is no gravitational degrees of freedom in three dimensions \footnote{Indeed, Einstein tensor and the curvature tensor are equivalent in 2+1 dimensions. This means that in source-free regions, the curvature tensor simply identifies with the empty space solution of Einstein's equations.}, which drastically simplifies general relativity (now an exactly solvable model \cite{witten19882+}). Within this framework, the geometry surrounding a point-particle is a conical singularity, the mismatch angle being proportional to the particle's mass. In other words, conical defects represent point particles coupled to gravity in 2+1 spacetimes. After Katanaev \cite{katanaev1992theory} first pointed out that the theory of linear disclinations was isomorphic to the 2+1-gravity, Kholodenko \cite{kholodenko2000use} used the apparatus of quadratic differentials to establish the connection between Deser, Jackiw and 't Hooft model and defects in liquid crystals. In essence, the existence of massive particles considered as field singularities is directly related to the topology of the underlying manifold (the Euler characteristic) and to the emergence of the induced saddles: this means that 2+1 Einstein's equations are strictly equivalent to the Poincaré-Hopf theorem (see section 5.2 in \cite{kholodenko2000use}), the Hopf quantization rule making the direct connection between particles masses $M_i$ and the defect topological charge \textit{m}, $4G M_i=m $ \cite{kholodenko2000useb}. 

\begin{figure}
\centering
\includegraphics[width=0.8\columnwidth]{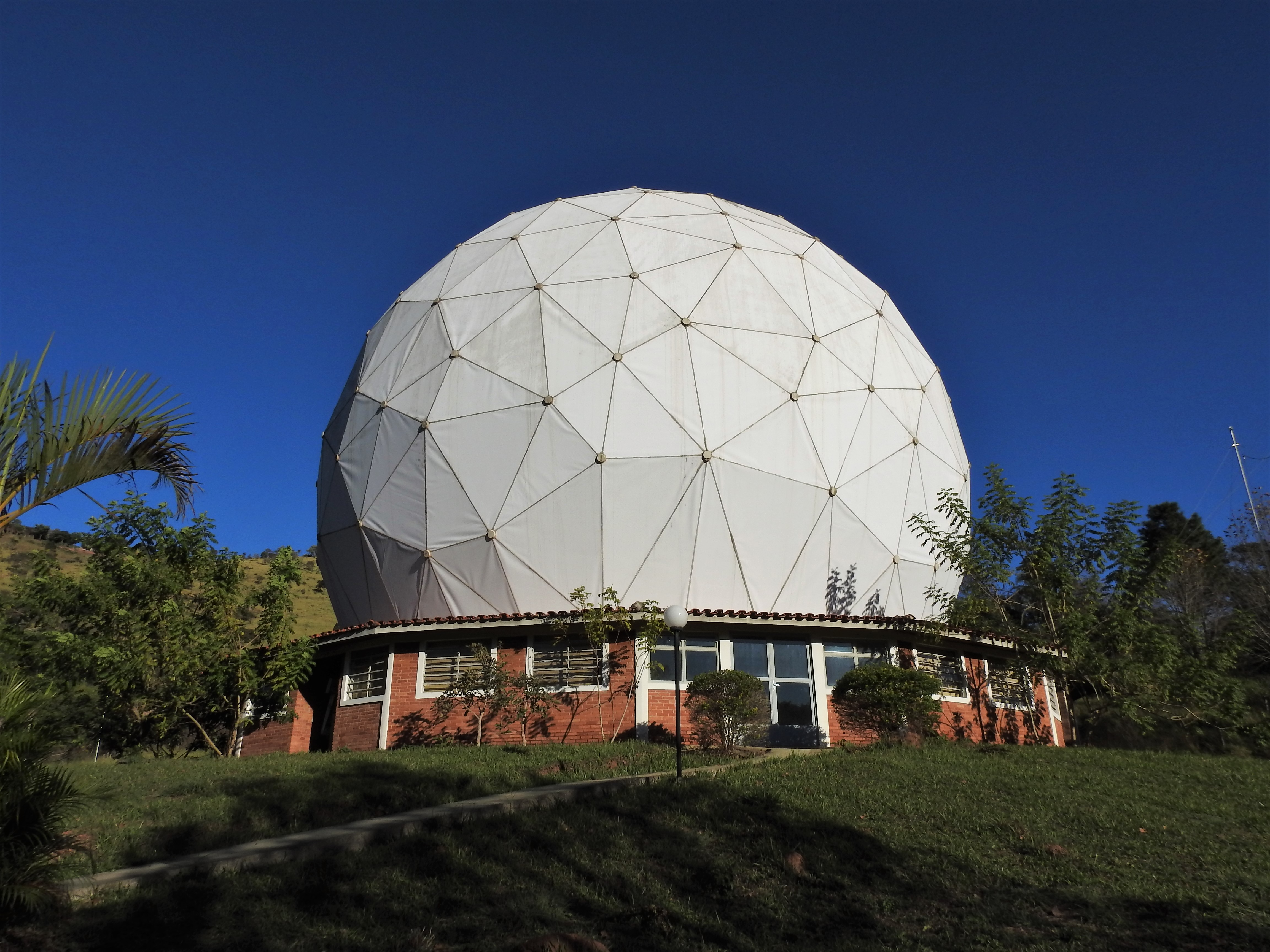}
\caption{Triangulation of a sphere at the Itapetinga radiotelescope (Brazil). Covering a curved surface by the triangles generates dihedral angles all around the sphere.}
\label{regge}
\end{figure}

The 2+1 gravity model can therefore be experimentally investigated from a network of parallel disclinations lines in a 3D nematic sample. Geometry of disclinations networks has been theoretically investigated in the literature, sometimes allowing for analytical expressions for the metric tensor \cite{letelier2001spacetime,fumeron2017geometrical,fumeron2020cmp}. Several authors have shown the possibility to design arrays of topological linear defects from photopatterning techniques \cite{wang2017artificial,guo2021photopatterned,sasaki2021general,nys2022nematic,harkai2022manipulation} and even to manipulate them \cite{liu2019flow,jiang2022active}. If this last point opens the possibility to emulate collisions between particles in the 2+1 model, it is even more interesting for the extension of the Deser, Jackiw and 't Hooft model to 3+1 dimensions \cite{t2008locally}: matter particles are represented by a gas of piecewise straight string segments that are likely to collide with a higher frequency. The strings display both positive and negative mass densities, i.e. they are associated to $\alpha<1$ and $\alpha>1$ Frank angles, which makes liquid-crystal-based experiments particularly promising to investigate such models. This model may also have deep connections with Regge calculus in quantum gravity, where the smooth curved spacetime is replaced by a piecewise-flat simplicial manifold. This is like the triangulation of a surface in 3D where the local curvature is described by the dihedral angle between adjacent triangles (the triangle is a 2D simplex). The effect of gluing the edges of the simplexes generate a network of cone-like singularities (Regge cones) which are analogs to wedge disclinations \cite{sorkin1975time,sorkin1981erratum} (see Fig. \ref{regge}).

\paragraph{Non-standard cosmology}

\begin{figure}
\centering
\includegraphics[width=1.0\columnwidth]{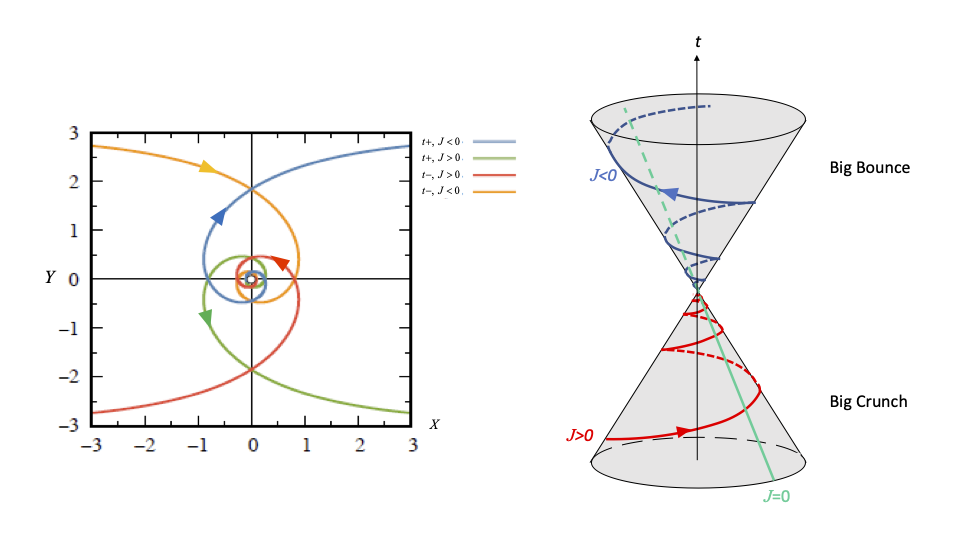}
\caption{Timelike geodesics with the radial time $t$ given in units of $t_0$ and $\kappa=1/3$. The blue and green (orange and red) lines are moving away (towards) from (to) the singularity. Furthermore, particles following the trajectories in the first (third) and second (fourth) quadrants are spinning clockwise (counterclockwise). The blank point at the origin is just to emphasize that the curves do not reach the singularity. Taken from \cite{figueiredo2017cosmology}. }
\label{doubleConefig}
\end{figure}

Cosmology at transplanckian scales is a thorny problem, both theoretically and of course observationally. For instance has the universe popped up from a unique singular event, the Big-Bang? And if so, how not to wonder what could have happened before it and how to design experiments to test these theories? Today, many high-energy physics theories such as quantum loop gravity and supertring theories entice the search for cyclic universe models, that is an endless repetition of big crunches followed by big bounces, along the same line of thought as the Stoics' concept of palingenesia. A safe transition has been proposed \cite{khoury2002big,steinhardt2002cosmic}, where the singularity is nothing more than the temporary collapse of a fifth dimension, the three space dimensions remain large and time keeps flowing smoothly. A toy model for the geometry of this transition is the compactified 2D Milne universe $\mathcal{M}_C$ \cite{horowitz1991singular,malkiewicz2006probing}. The Milne universe metric is given by
\begin{equation}
	ds^2=-dt^2+t^2 d\chi^2+t^2 \sinh^2\chi \left(d\theta^2+\sin^2\theta d\phi^2 \right) \label{RW_milne}
\end{equation}
and it was proposed by E.A. Milne in $1933$. It represents a homogeneous, isotropic and expanding model for the universe with a negative curvature. In order to compactify the Milne universe on hypersurfaces of fixed solid angle, let the variable $\chi$ acquire some period denoted as $2\pi \kappa$: here, $0<\kappa \in \mathbb{R}^1$ is a constant parameter for compactifications. After reparametrization, the line element corresponding to the compactified Milne universe finally writes as
\begin{equation}
	ds^2=-dt^2+\kappa^2t^2d\phi^2
    \label{milne_metric}
\end{equation}
\noindent where $t \in \mathbb{R}^1$. As can be seen from the disclination line element (\ref{wedge-discli}), the presence of $\kappa^2$ in the above metric indicates a conical singularity of the curvature at the origin (see Fig. \ref{doubleConefig}).

The passage through the initial singularity has several unusual features \cite{fumeron2015generation,figueiredo2017cosmology}: the singularity acts as a filter for classical particles and a phase-eraser for quantum ones. The timelike geodesics of (\ref{milne_metric}) reveal that depending on their angular momentum $J$, particles have two ways of crossing the singularity: 1) For non-vanishing $J$, a two-step dynamics consisting in an inward stable motion before the singularity, followed by outward stable motion on the other side, with a memory loss of the particle kinematical properties (quantum mechanically, this effect simply comes from strong oscillations of the phase of the wave function at the singularity. 2) For $J=0$, a one-step dynamics consisting in a straight line through the collapse: yet such trajectories are very unstable since small perturbations in the value of $J$ causes large deviations on the trajectories.

Probing how particles behave in $\mathcal{M}_C$ can be tested in the laboratory from hyperbolic liquid crystal metamaterials (HLCM): this means that the permittivity along the director axis $\varepsilon_{\parallel}<0$ and the permittivity perpendicular to the director axis $\varepsilon_{\perp}>0$ are of opposite sign \cite{figueiredo2016modeling}. Such media can be made from a host nematic liquid crystal that includes an admixture of metallic nanorods \cite{xiang2012liquid} or coated core-shell nanospheres \cite{pawlik2014liquid}. To retrieve the Kleinian double-cone geometry, the HCLM must be endowed with an hyperbolic disclination: the line element writes as $ds^2=\varepsilon_{\perp}d\rho^2-\varepsilon_{\parallel}\rho^2d\phi^2+\varepsilon_{\perp} dz^2$, which after a rescaling becomes $ds^2=-\gamma^2r^2\phi^2+dr^2+dz^2$. This line element is relevant only by the extraordinary modes and for radial injection conditions (planar trajectories $z=C^{st}$), the geometry experienced by extraordinary rays is perfectly identical to that of the compactified Milne universe.

A stable configuration for the director field may be obtained from a cylindrical shell of HLCM with homeotropic anchoring at the boundaries. In the geometrical optics limit, extraordinary light paths turn out to be Poinsot's spirals as for the compactified Milne universe. The practical realization sets limits to the efficiency of such analog device for the classical particles. First, the analysis holds only within a limited frequency bandwidth due to the resonant nature of the used core-shell spheres. Second, as previous phenomena concern the extraordinary mode, an efficient optical absorber should include a filter to shut off the ordinary wave. Finally, it should be noticed that the present model concerns optics inside a bulk hyperbolic material: to design a perfect optical analog, the hyperbolic medium must be impedance matched to avoid sizable reflections at the interfaces. The analogy was also extended to quantum particles by investigating light in the scalar wave approximation in the same device.

\section{Tayloring transport with liquid crystals: defect-engineered materials}

\subsection{Acoustics}

\paragraph{Ballistic guiding} The geometric description of linear defects revealed wedge disclinations carry curvature along their axis: a positive Frank angle corresponds to a conical geometry that focuses incoming light rays, in similar fashion to what happens with a converging lens. Conversely, a negative Frank angle corresponds to a saddle-like geometry that scatters geodesics, in the same fashion as a diverging lens. It is worth noticing that the effect of a disclination does not limit to the eikonal approximation, as it also diffracts incoming waves: from the present geometric approach, computing the differential scattering cross section of a wedge disclination showed good agreements with theoretical results obtained by Grandjean from standard acoustics \cite{pereira2011diffraction,pereira2013metric}.

 \begin{figure}
\centering \includegraphics[width=1\linewidth]{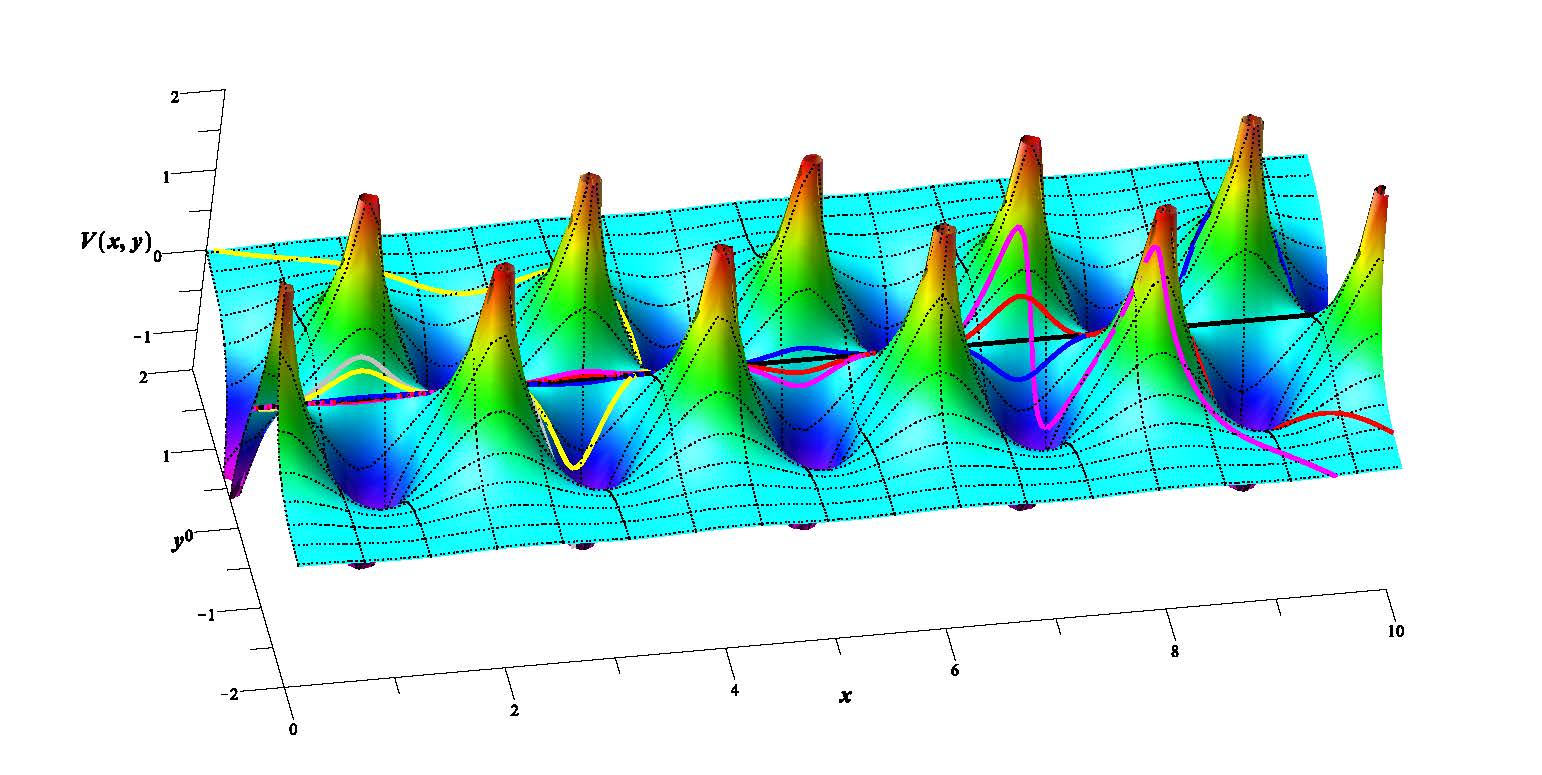} 
\caption{Geodesics of the channel of defects in the potential landscape. Taken from \cite{fumeron2015optics}.}
\label{Fig3dGeodesics} 
\end{figure}

\begin{figure}
\centering \includegraphics[width=1\linewidth]{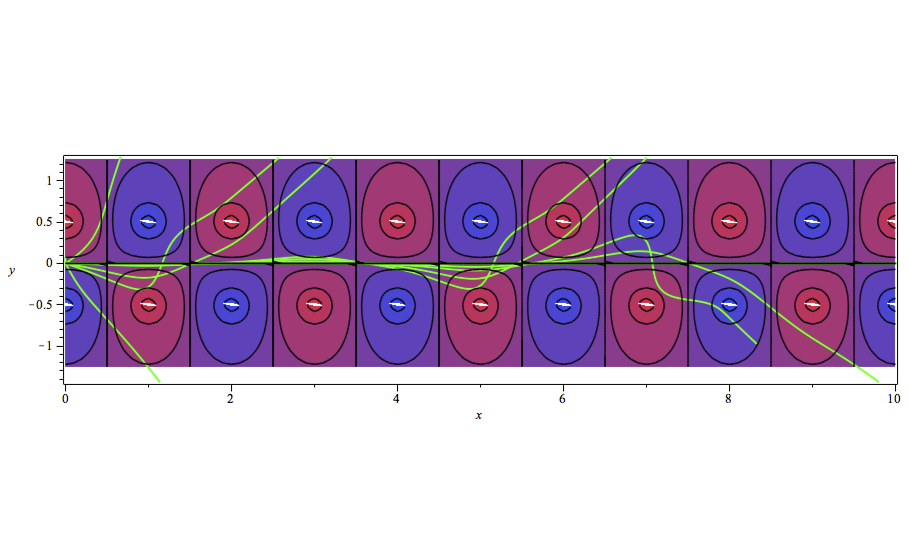} \vspace{-1cm}
 \caption{Different geodesics, shot from the origin, in the channel of disclinations geometry. The positive disclinations correspond to red contours while negative disclinations correspond to blue contours. Depending on the shooting angle, the propagation of phonons may be guided by the street of topological defects. Taken from \cite{fumeron2015optics}.}
\label{FigGeodesics1} 
\end{figure}

The effect of a single wedge disclination on rays being clarified, one can legitimately wonder if a well-suited arrangement of such defects can be used to taylor the propagation of sound (we recall that photopatterning techniques now allow for the practical realization of almost any kind of arrays). In \cite{fumeron2017geometrical}, a channel of disclination dipoles was considered, consisting in two infinite rows made of alternate disclinations separated by distance $2a$ (a kind of von K\'arm\'an alley), the distance between the rows being $2b$. The positive disclinations are at points located at $(na,(-1)^{n}b),n\in\mathbb{Z}$, while negative disclinations have coordinates $(na,(-1)^{(n+1)}b),n\in\mathbb{Z}$. As always done in geometric models, the bulk medium is considered in the continuum limit (i.e. limit of a vanishing lattice spacing). The corresponding background geometry writes as
\begin{equation}
ds^{2}=-c^{2}dt^{2}+e^{-4V(x,y)}\left(dx^{2}+dy^{2}\right)+dz^{2}=g_{\mu\nu}dx^{\mu}dx^{\nu} \label{Lmetric}
\end{equation}
where $c$ is the local speed of wave packet and $V$ is the acoustic potential given by \cite{fumeron2017geometrical}:
\begin{eqnarray}
V(x,y)=\;\;\;\; \;\;\;\; \;\;\;\; \;\;\;\; \;\;\;\; \;\;\;\;\;\;\;\;\;\; \;\;\;\; \;\;\;\; \;\;\;\; \;\;\;\; \;\;\;\;\;\; \;\;\;\; \;\;\;\; \;\;\;\; \;\;\;\; \;\;\;\; \;\;\;\;\;\;\;\;\;\; \;\;\;\; \;\;\;\; \;\;\;\; \;\;\;\; \;\;\;\;\;\; \nonumber \\
\frac{\lvert F\rvert}{4\pi}\ln\left[\left(\frac{\cosh^{2}\left(\frac{\pi}{2a}(y-b)\right)-\cos^{2}\left(\frac{\pi x}{2a}\right)}{\cosh^{2}\left(\frac{\pi}{2a}(y-b)\right)-\sin^{2}\left(\frac{\pi x}{2a}\right)}\right)\left(\frac{\cosh^{2}\left(\frac{\pi}{2a}(y+b)\right)-\sin^{2}\left(\frac{\pi x}{2a}\right)}{\cosh^{2}\left(\frac{\pi}{2a}(y+b)\right)-\cos^{2}\left(\frac{\pi x}{2a}\right)}\right)\right]  \nonumber \\ 
\label{fhat}
\end{eqnarray}
where $\lvert F \rvert$ is the absolute value of the Frank angle that characterizes the $\pm F$ defects.
The sound paths are attracted by the positive defects while repelled by the negative ones (see Fig. \ref{Fig3dGeodesics} and \ref{FigGeodesics1}). Adjusting the defect strengths along with parameters governing the geometry of a cell (namely $a$,$b$) opens the possibility to taylor material properties of the sheets, but this will only be achieved numerically considering the complexity of analytical expressions. Sound paths are yet very sensitive to the shooting angle. Hence, a thorough optimization of the distribution of defects deserves an additional treatment of chaos involving the statistical tools of dynamic hamiltonian systems.

\paragraph{Acoustic rectification} Differential geometry models for acoustics originates from \cite{fischer2002riemannian} where vorticity effects in isotropic fluids were investigated. In nematics, for a planar horizontal configuration, the acoustic metric experienced by the extraordinary mode is formally identical to (\ref{general-discli}) with the substitutions $\varepsilon_{\bot}\leftrightarrow \rho v^2/C_{33}$ and $\varepsilon_{\Vert}\leftrightarrow\rho v^2/C_{11}$, where $\rho$ is the mass density, $v$ is the velocity of sound in the isotropic phase, $C_{33}$ and $C_{11}$ are elastic constants respectively along the director's direction and in directions orthogonal to it. When the nematic medium is confined inside a capillary tube (radius $R$) with homeotropic boundary conditions, the director field tends to be radially oriented everywhere but on the central axis where an orientational singularity lies, but to reduce the elastic energy of this configuration, the director escapes in the third dimension \cite{cladis1972non,meyer1978observation} and the nematic relaxes into a funnel-shaped configuration, known as the escaped radial disclination (ERD) where the delta-distributed Ricci scalar becomes an extended smooth one \cite{fumeron2016retrieving}.

\begin{figure}[H]
\centering \includegraphics[width=1.1\linewidth]{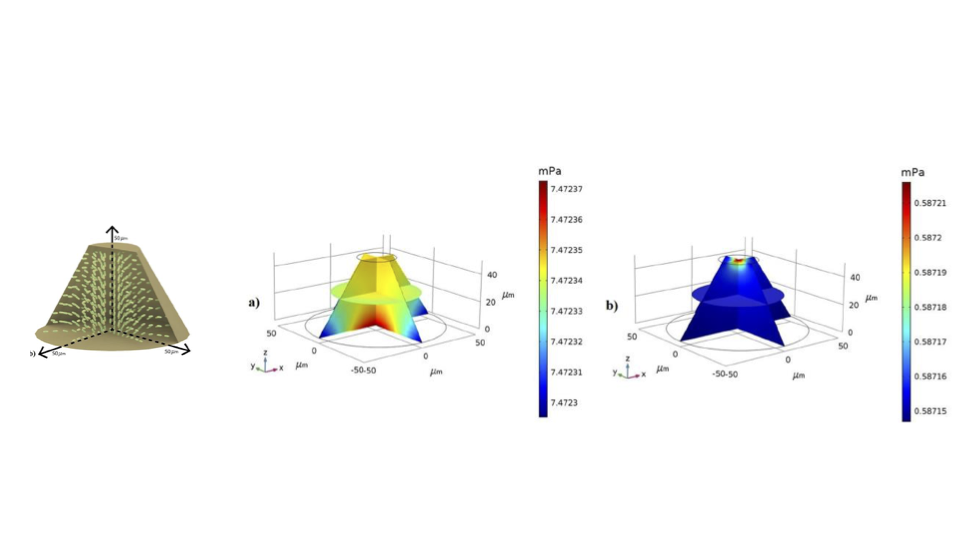} \vspace{-1cm}
 \caption{\textit{Left}: Conical frustum with an ERD. \textit{Middle}: Pressure field for the incoming waves from bottom to top. \textit{Right}:Pressure field for the incoming waves from top to bottom. Taken from \cite{viana2019high}.}
\label{acoust-diode} 
\end{figure}

In general, rectification effects come from an asymmetry of the system in the direction along which the transport phenomenon occurs. Usually, this is generally achieved by relying on gradients of physical properties (e.g. pore density \cite{criado2013thermal}, distribution of compositional defects \cite{dettori2016thermal}...), asymmetric geometries \cite{sawaki2011thermal,zhang2016transition}... all examples of situations implying hard and non-flexible systems. Liquid crystals naturally provide a stable but flexible configuration corresponding to such asymmetry, the ERD. In the spirit of a low-cost soft-matter-based solution, satisfying levels of acoustic rectification have been obtained by combining the asymmetry of the ERD configuration to that of a container consisting in conical frustum of varying radius $R(z)$ \cite{silva2018high}. The inner surface prepared to produce the desired anchoring angle\footnote{Whereas homeotropic anchoring of nematics on flat surfaces is well mastered \cite{oswald2005nematic}, anchoring nematics with an arbitrary angle on a curved surface is generally not trivial and it is particularly sensitive to the saddle-splay constant $K_{24}$ \cite{crawford1992surface}.}. The anchoring angle adapted to maintain the ERD configuration. $\alpha$ depends on the surface geometry (radius), on the liquid crystal nature (elastic constant $K$, saddle-play constant $K_{24}$) and on the surface treatment. 
 Reference \cite{silva2018high} considered acoustic waves in a frequency range between $20$ Hz and $20$ kHz (the average human audible range) propagating in 5CB for the ERD configuration. The rectification parameter used to estimate the acoustic device’s efficiency is the percent standard deviation of the lowest variation on the acoustic intensity
\begin{equation}
    \text{Acoustic rectification}(\%)=\displaystyle\left\lvert \frac{\Delta I_{bt}-\Delta I_{tb}}{\text{min}\left(\Delta I_{bt},\Delta I_{tb}\right)}\right\rvert\times 100
\end{equation}
where $\Delta I_{bt}=I_t-I_b$ is the acoustic intensity variation if the wave comes from the bottom to the top of the conical frustum and $\Delta I_{tb}=I_b-I_t$ is the analogous variation for the counterpropagating case. Numerical simulations show a rectification effect for a longitudinal plane wave propagating along the conical frustum axis (see Figure \ref{acoust-diode}). The optimization of the device parameters (geometry of the conical frustum, anchoring conditions the rectification effects...) allows to reach rectification levels up to $1300\%$ for a continuous frequency bandwidth \cite{viana2019high}.

\subsection{Optics} 

\paragraph{Waveguiding and light concentration}

Manipulation of light has become a major issue in a large number of applications ranging from solar energy harvesting to optical sensing. The beam steering technique relies on a modulation of refractive indices to guide light in a given direction (\cite{fray1975large,sasaki1979liquid,masuda1997liquid,2005Apter}). More recently, the possibility to continuously deflect in arbitrary directions and/or to simultaneously focus/defocus an incoming light beam has been demonstrated from multiple stacked nematic liquid crystal cells—building blocks \cite{mur2022controllable}. Another possibility to guide light beams is to use optical waveguides with nematic cores in radial escaped disclination configurations: the defocusing due to natural diffraction is compensated by the converging lens effect resulting form negative birefringence  \cite{Lin1991,Lin1992,Ravnik2016,vcanvcula2015nematic,brasselet2021singular}. 

\begin{figure}[H]
\centering
\includegraphics[scale=0.35]{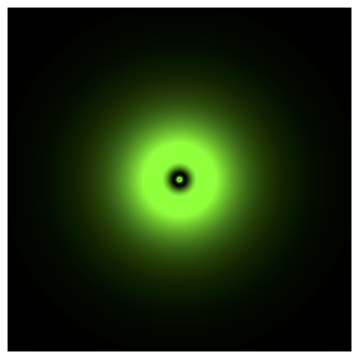}
\includegraphics[scale=0.35]{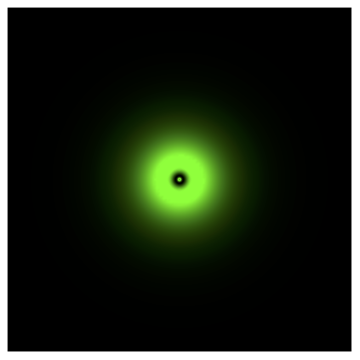}
\\
\includegraphics[scale=0.35]{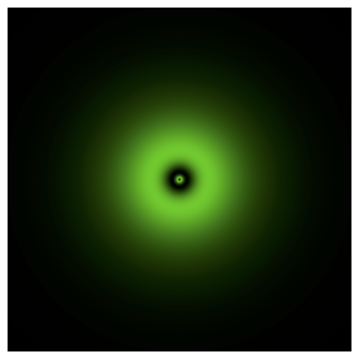}
\includegraphics[scale=0.35]{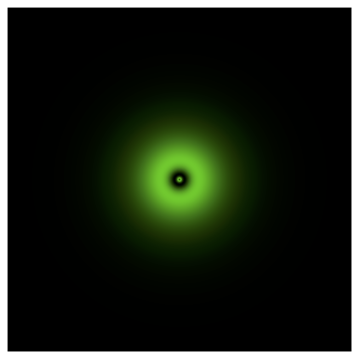}
\caption{Examples of intensity profiles, representing the transverse field distributions concentrated in the vicinity of the hyperbolic defect. Taken from \cite{azevedo2018optical}.  
}
\label{intensity}
\end{figure}

Light focusing has long been suspected of suffering severe limitations, the Rayleigh criterion forbidding beam sizes below half of a wavelength. Recently, the advent of metamaterials  provides new hopes for overcoming the diffraction limit using superlenses \cite{zhangNature2008}. Another promising possibility is to use an hyperbolic liquid crystal metamaterial (HLCM), obtained from an admixture of metallic nanoobjects to nematics \footnote{The losses due to metallic components do not jeopardize the performances of the device, as they might be offset by using gain media as pointed in \cite{xiang2012liquid} (highly doped oxides with lower dissipation levels have also been considered ). The low-loss limit for metamaterials can reached by working within the terahertz waveband.}. As seen in the previous section, the permittivity along the director axis $\varepsilon_{\parallel}$ and the permittivity perpendicular to the director axis $\varepsilon_{\perp}$ have opposite signs. For an orthoradial director field, the effective metric writes as $g_{ij}=\hbox{diag}\ \!(1,-\alpha^2 \rho^2,1)$ and in the eikonal approximation, the light paths are the aforementioned Poinsot spirals: the hyperbolic defect behaves as a sink for light paths \cite{azevedo2018optical}. The smaller the value of $\alpha$, the stronger is the spiraling behavior, $1/\alpha$ corresponding to the defect vorticity. 
Concentration of light by an hyperbolic disclination extends beyond the geometrical optics limit. In the scalar wave approximation, the complex amplitude $\Phi$ of the wave is governed by the generalized form of the d'Alembert equation, involving the Laplace-Beltrami operator instead of the ordinary Laplace operator. The wave equation writes as a modified Bessel differential equation of imaginary order $i\ell/\alpha$ and the solutions are linear combinations of the modified Bessel functions of first and second kind, respectively. The intensity distribution for the propagating fields shows 1) that the electromagnetic field concentrates along the axis of the device, and 2) that the bigger the value of the frequency, the smaller the light rings are (see Fig. \ref{intensity}). 

\paragraph{Optical vorticity} 

Active beam shaping has also emerged as a major trend in modern optics. The idea is to taylor the amplitude, the phase or the polarization of an optical wavefront from real-time driven systems that ideally must be compact enough, highly-flexible but yet have low manufacturing costs. Ought to their high response functions, liquid crystal-based devices naturally fulfill all these requirements and have emerged as promising low-cost and easy-to-manufacture alternatives to MEMS, moving opto-mechanical devices and photonic crystals \cite{lavrentovich2011liquid,shang2020active}. 

In nematics, the orbital angular momentum carried by light beams can be tuned from $q$-plates. A ${q}$-plate consists in an inhomogeneous liquid crystal cell endowed with a disclination of topological charge \textit{q} (for details regarding the dielectric tensor, see \cite{vaveliuk2010}). When an electromagnetic wave propagates inside the medium, its components  acquire designed phase shifts (Pancharatnam-Berry phase) that trigger spin-to-angular momentum conversions. This results in light beams displaying optical vorticity, i.e. the wavefronts are helical and the intensities profiles distribute in doghnut-like shapes \cite{marrucci2006pancharatnam,marrucci2007rotating,marrucci2008generation,slussarenko2011tunable,marrucci2013q,rubano2019q}. \cite{vcanvcula2014generation} demonstrated from FDTD simulations that disclination lines can transform the state of polarization of beams propagating along their axis. Umbilics ending disclination lines are also identified as structures generating optical vortex arrays at predetermined wavelength \cite{brasselet2012tunable}. 

Nematics are not the only contenders in the contest for optical vorticity. In Ref.~\cite{voloschenko2000optical} the Raman-Nath diffraction was used to generate optical vortices from edge dislocations in the stripe pattern of a cholesteric liquid crystal (cholesterics are chiral nematics for which the director whirls around a well-defined direction). Cholesterics were also considered in \cite{kobashi2016polychromatic,kobashi2016polychromatic}, where it has been theoretically and experimentally demonstrated that optical vortices were generated from a Bragg-reflection-based device, therefore likely to operate at multiple wavelengths. Recently, charged particles crossing a cholesteric plate were reported to radiate purely twisted photons \cite{bogdanov2021generation}. Chiral nematics are also likely to generate defect textures of a more complex kind than that optical dislocations \cite{wu2022hopfions}. Beside cholesterics, smectics also showed their potentialities for optical vorticity. \cite{son2014optical} produced an optical vortex from focal conic domains, whereas in \cite{fumeron2015generation}, screw dislocations in smectics were shown to imprint their torsion onto wavefronts (see also \cite{bazhenov1990laser,salamon2018tunable} for a solid-state-oriented context).

\subsection{Heat transfer}

\paragraph{Principles of thermal design} Manipulation of heat flux raises intensive research efforts because of the abundant wealth of potential applications including thermal shielding or stealth of objects, concentrated photovoltaics or thermal information processing (heat-flux modulators, thermal diodes, thermal transistors and thermal memories). These prospects come from the possibility of designing energy paths in a fashion similar to that of light in transformation optics. To do so, the first step is to understand the main peculiarities of heat transfer in the presence of a non-Euclidean geometry. Generally speaking, diffusion of a passive scalar (for instance the temperature field) can be seen as a collection of Markov processes obeying the stochastic Fokker-Planck equation. In the case of Brownian motion, the Fokker-Planck equation reduces to the well-known parabolic heat equation \cite{pavliotis2014stochastic}. When considering diffusion processes in the presence of a non-Euclidean space, the problem is addressed, as already discussed, by replacing the Laplace operator with the Laplace-Betrami operator $\Delta_{LB}$ \cite{smerlak2012tailoring}: 
\begin{equation}
\frac{\partial T}{\partial t}=D \Delta_{LB} T . \label{Eckart}
\end{equation}
Here, $D$ is the diffusivity and its value depends on the material properties. Ought to the form of the metric of a wedge disclination, heat conduction locally occurs as in a monoclinic-like crystal with no internal source \cite{fumeron2018thermal}: the heat flux vectors are no longer perpendicular to the isothermal surfaces, which are bent depending on the value of the Frank angle. In other words, disclinations in nematics generate thermal lensing effects.

\begin{figure}[!ht]
\begin{center}
\includegraphics[height=14cm]{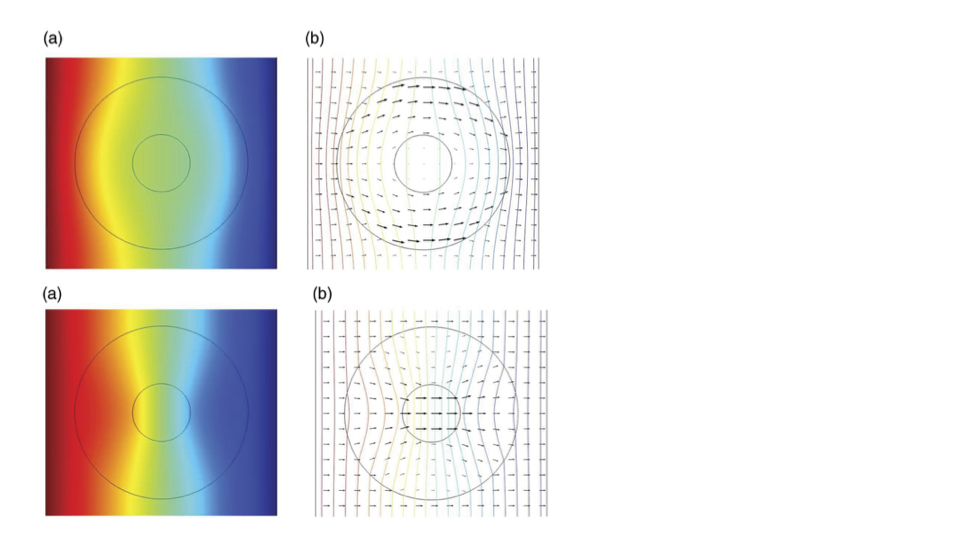} \caption{\textit{Top}: Temperature field (a) and heat flux field (b) for a radial director field (homeotropic anchoring) and $\alpha=\sqrt{C_{33}/C_{11}}=0.5$ \textit{Bottom:} Temperature field (a) and heat flux field (b) for an orthoradial director field (parallel anchoring) such that $\alpha=\sqrt{C_{11}/C_{33}}=2$. Taken from \cite{fumeron2014principles}}\label{device-heat-PRE}
\end{center}
\end{figure}

Once the basic effects of single line defects are understood, the next step is to taylor them to guide heat. To do so, let us consider a hollow cylinder, inside which there is the core region where one aims at controlling the conductive heat flux. The cylinder is inserted inside a conducting solid sandwiched between two heated vertical plates. The host material consists of a homogenous isotropic medium, whereas the intermediate thick cylinder consists of a nematic liquid crystal in a disclination-like configuration (no disclination core). For thermal management, mesophases with low melting and high clearing temperatures are required: a range of about 100 K can be reached by using eutectic liquid crystal mixtures (or ``guest-host systems''). Numerical simulations  \cite{fumeron2014principles} confirm the possibility of a strong heat guiding phenomenon: depending on the value of elastic constants $C_{33}$ (along the director) and $C_{11}$ (along any direction perpendicular to the director), the device can either cloak the core region from the heat flux or concentrate heat there (see Fig. \ref{device-heat-PRE}). Switching from the concentrator to the cloaking device is achieved by an electric-field-driven bistable anchoring with dye-doped mematics (sufficiently high values of the electrical potential difference between the two sides of the hollow cylinder were indeed shown to induce stable anchoring transitions between homeotropic and parallel states \cite{kim2010heat}). To avoid thermoconvective instabilities in the annulus domain, the device must be thin enough and the heat flux and temperature levels must be moderate. For instance, using 5CB and MBBA, if the temperature is about a few tens of degrees (thermoelectric applications) and the external radius of a few centimeters, the device handles heat flux that typically varies from $5\ \!{W/\rm m}^2$ (repeller) to $103\ \!{\rm  W/m}^2$ (concentrator).

\paragraph{Thermal diodes} The previous study is now refined in order to investigate thermal rectification. Thermal rectification is a very active subject in nanoscience and solid-state physics, as testified by the abundant litterature dealing with this subject (extensive reviews treating thermal
rectification can be found in \cite{wehmeyer2017thermal,wong2021review}. More seldomly is heat conduction rectification considered from soft-matter-based devices. In liquid crystals, the macroscopic thermal properties of the nematic phase depend on temperature according to Haller's approximation \cite{haller1975thermodynamic}:
\begin{eqnarray}
\lambda_{\parallel}(T)&=&\lambda_0+\lambda_1\times\left(T-T_{NI}\right)+\lambda_{1\parallel}\times\left(T-T_{NI}\right)^{\alpha_{\parallel}} \\
\lambda_{\perp}(T)&=&\lambda_0+\lambda_1\times\left(T-T_{NI}\right)+\lambda_{1\perp}\times\left(T-T_{NI}\right)^{\alpha_{\perp}}
\end{eqnarray}
where $\lambda_0$, $\lambda_1$, $\lambda_{1\parallel}$, $\lambda_{1\perp}$, $\alpha_{\parallel}$ and $\alpha_{\perp}$ are material-depending constants, whereas $T_{NI}$ and $T_C$ are, respectively, the nematic-to-isotropic temperature and the clearing-point temperature of the liquid crystal. As discussed before, liquid crystals naturally provide a stable and flexible configuration corresponding to such asymmetry, the ERD, which already turns out to provide high levels of acoustic rectification. To achieve high rectification levels, the same conical frustum of varying radius $R(z)$ with anchoring conditions can be used. In analogy with the acoustic case discussed earlier, the rectification parameter used to estimate the thermal diode efficiency can be defined as
\begin{equation}
    \text{Thermal rectification}(\%)=\displaystyle\left\lvert \frac{\Delta T_i-\Delta T_d}{\Delta T_d}\right\rvert\times 100
\end{equation}
where $\Delta T_d=T_{d,h}-T_0$ is the difference between  $T_{d,h}$, the high
temperature on one base produced by the heat pumped
in the cylinder when working in the direct setup (i.e. when the heat
is flowing from the narrow region
to the wider one, i.e. the $-z$ direction),
and $T_0$, the temperature at the other base, which is also the
initial temperature. Similarly, $\Delta T_i=T_{i,h}-T_0$ when
working in the inverse setup. 

Numerical simulations show thermal rectification rates around $1266\%$ \cite{melo2016thermal}. On the shape parameters, alterations on the ratio $R_r \in [0,28;0,75]$ produced a percentage variation on the thermal rectification around $1273\%$, while modifications of the height $h \in [50;75]$ $\mu$m and on the larger radius $R_{l} \in [50; 70]$ $\mu$m produced percentage changes lower than $5\%$. This indicates that the anisotropy of the conical frustum tube has a strong influence on the rectification. Other non-geometrical parameters such as the anchoring angle (in the range $[0;90^{\circ}]$) and the inward pumped heat flux (in the range $[5;10]$ kW/m$^2$) give percentage variations on the rectification around, respectively, $3,8$ and $1,7\%$. Such characteristics enable this improved thermal diode to be miniaturized, applied on well-determined areas, while robust against variations of the inward pumped heat flux.

\begin{figure}
\begin{center}
\includegraphics[width=1\linewidth]{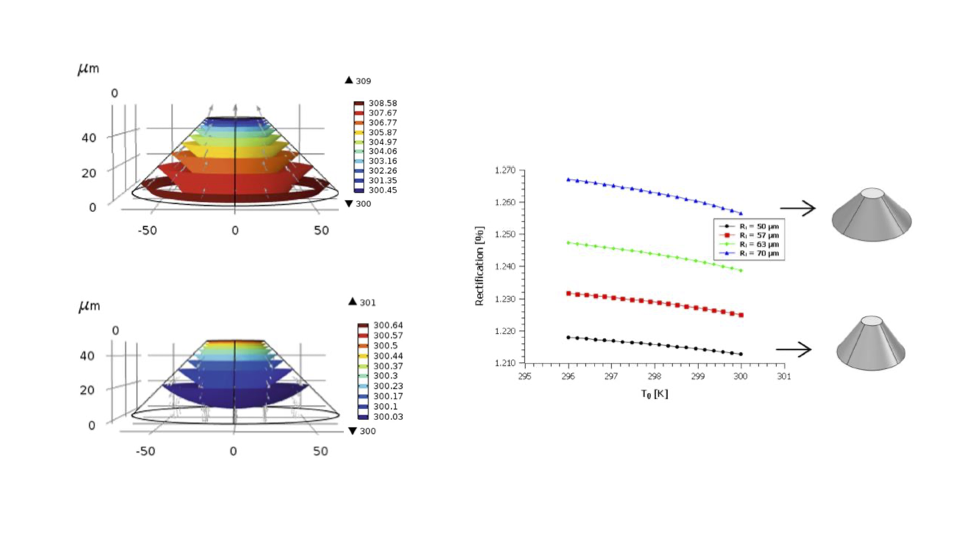} \caption{\textit{Left}: Isothermal surfaces of a liquid crystalline thermal diode in (up) inverse thermal setup and (bottom) direct thermal setup. The frustum diode has larger radius $R_l = 70$ $\mu$m, ratio between the radii is $R_r = R_{sm}/R_{l}= 0,28$, the height is $h = 50$ $\mu$m, anchoring is $60^{\circ}$ and the inward heat flux is $Q=5\:$kW/$m^2$ on the base with the higher temperature and $T_0$=296 K. \textit{Right:} Rectification rate versus temperature $T_0$ of the base for different larger radii $R_l$. Taken from \cite{silva2018high}. }\label{device-heat}
\end{center}
\end{figure}

 The identical forms of the geometry experienced by light and by sound strongly suggests that devices using liquid crystals may be used to manipulate simultaneously  optical and thermodynamical transport. Indeed, Ref. \cite{barros2018concurrent} reports the control of both electromagnetic propagation and heat flow by a liquid crystal device similar to the one depicted in Fig.  \ref{device-heat-PRE}, while Ref. \cite{santos2018simultaneous} uses an escaped disclination configuration to rectify at the same time both heat and light, thus a thermo-optical diode.

\section{Conclusion and perspectives}

Since their early discovery in the XIX$^{th}$ century, liquid crystals have been the magic bullet in physics and engineering. Being in-between anisotropic solids and isotropic fluids, they extended our conception of condensed matter to an area where geometry and topology can be almost as useful as in General Relativity. An orientationally ordered fluid, as the nematic liquid crystal, is a vivid  representation of a Riemannian manifold where the director field can be associated to a local vector basis (triad or \emph{dreibein}). The relative rotation of the director/triad associated to neighboring points indicates the presence of curvature. Boundary conditions like vessel shape, immersed objects, anchoring angle, etc., impose restrictions to the effective geometry whose eventual incompatibility with the nematic order (ground state or zero curvature everywhere) leads to the appearance of topological defects which accommodate the incompatibilities. 

This geometric view of the elastic distortions in the NLC is complemented by the optical effective geometry that appears naturally by comparing Fermat's law of least time to the geodesic variational principle. Similar effective geometries can be obtained for acoustics and heat transport as well. In all these cases the defects, besides being the consequence of the topology (boundary conditions), are the source of the geometry. One might then say that, as soon as there is real or effective curved geometry to describe a physical system with orientational order, one can expect defects. And, if there is curved geometry, one can relate NLC to gravitation and cosmology. In this article we reviewed, not only this relationship, but also the physical applications obtained with the help of the geometric tools. Many open problems both in gravitation and cosmology and in NLC certainly may benefit from the analogies derived by the (partially) common geometry. For instance, the experimental knowledge about the inner structure of disclinations may be an inspiration for cosmic string core models.
Active matter, being a dynamic medium, may be described by time-depending metrics. Furthermore, being a dissipative medium, active matter (or its effective geometry) might obey a geometric flow like Ricci's \cite{chow2008ricci} in its way to the equilibrium.


\vspace{0.5cm}
\begin{acknowledgements}
\noindent 
For the purpose of Open Access, a CC-BY public copyright licence, \href{https://creativecommons.org/licenses/by/4.0/}{\includegraphics[width=0.12\columnwidth, angle=0]{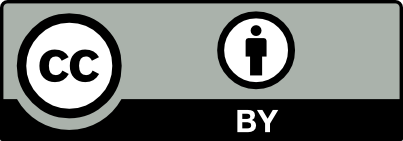}},
has been applied by the authors to the present document and will be applied to all subsequent versions up to the Author Accepted Manuscript arising from this submission. 
\end{acknowledgements}

\bibliographystyle{unsrt}
\bibliography{references}

\end{document}